\documentclass[preprint,twocolumn,10pt,a4paper,byrevtex,unsortedaddress,superscriptaddress,prb]{revtex4}
\usepackage{float}
\usepackage{amssymb}
\usepackage{natbib}
\usepackage{graphicx}
\usepackage{hyperref}
\usepackage{subfigure}
\usepackage{amsmath}
\usepackage{color}
\usepackage{appendix}
\usepackage{tabularx}
\usepackage{siunitx}
\usepackage{doi}%<----------
\usepackage{gensymb}
\usepackage{xfrac}

\definecolor{DarkGreen}{rgb}{0,0.70,0}

\definecolor{DarkBlue}{rgb}{0,0,0.8}

\definecolor{purple}{rgb}{0.5,0,1}

\definecolor{orange}{rgb}{1,0.5,0}

\definecolor{Pink}{rgb}{1,0,1}

\definecolor{Red}{rgb}{1,0,0}

\newcommand{\lt}{\ensuremath <}

\DeclareUnicodeCharacter{2009}{\,} 
\DeclareUnicodeCharacter{0306}{z}

\begin{document}

\title{Superconducting spintronics with electron symmetry filtering and interfacial spin-orbit coupling}

\author{Pablo Tuero\textsuperscript{\textdagger}}
\affiliation{Dpto. Fisica de la Materia Condensada C-03, Universidad Autonoma de Madrid, Madrid 28049, Spain}

\author{C\'esar Gonz\'alez-Ruano\textsuperscript{\textdagger}}
\affiliation{Dpto. Fisica de la Materia Condensada C-03, Universidad Autonoma de Madrid, Madrid 28049, Spain}
\affiliation{Department of Electrical Engineering and Institute for Research in Technology, ICAI School of Engineering, Comillas Pontifical University, C/Alberto Aguilera 25, 28015 Madrid, Spain}

%\author{Chenghao Shen}
%\affiliation{Department of Physics, University at Buffalo, State University of New York, Buffalo, NY 14260, USA}
%\author{Jong E. Han}
%\affiliation{Department of Physics, University at Buffalo, State University of New York, Buffalo, NY 14260, USA}
\author{Igor \v{Z}uti\'c}
\affiliation{Department of Physics, University at Buffalo, State University of New York, Buffalo, NY 14260, USA}

\author{Yuan Lu}
\affiliation{Universit\'e de Lorraine, CNRS, Institut Jean Lamour, F-54000 Nancy, France}

\author{Coriolan Tiusan}
\affiliation{Department of Solid State Physics and Advanced Technologies, Faculty of Physics, Babes- Bolyai University, Cluj-Napoca, 400114 Romania}

\author{Farkhad G. Aliev*}
\affiliation{Dpto. Fisica de la Materia Condensada C-03, Instituto Nicolas Cabrera (INC) and Condensed Matter Physics Institute (IFIMAC), Universidad Autonoma de Madrid, Madrid 28049, Spain}
\date{\today }

\begin{abstract}

Over the recent years, the crossroads of magnetism and superconductivity have led to the emerging field of superconducting spintronics. A cornerstone of this venture is the generation of equal-spin triplet Cooper pairs in superconductor–ferromagnet hybrids, enabling long-range spin-polarized supercurrents and magnetic control over superconducting quantum states for the development of energy-efficient cryogenic devices. Until now, nearly all superconducting spintronic devices have relied on direct interfaces between superconductors and ferromagnets, since it was believed that an insulating barrier would decouple spin and charge transport. This assumption, however, appears to be invalid when a thin spin- and orbit-filtering barrier couples an epitaxial ferromagnet and a superconductor. Symmetry filtering plays a crucial role in enhancing giant tunneling magnetoresistance (TMR) by selectively allowing specific electronic states to tunnel through the barrier. Such a mechanism is key for high-performance spintronic devices like magnetic random access memory, magnetic sensors or spin-light emitting diodes. On the other hand, spin-orbit coupling (SOC) is a central mechanism for perpendicular magnetic anisotropy in spintronics. Recently, it has become clear that SOC is crucial in mediating the interactions in heterostructures combining superconductors and ferromagnets, otherwise antagonistic materials where exotic interfacial quantum phenomena have been discovered over the last decade. 
Building on recent advances in studies of various V/MgO/Fe(100)-based systems, this manuscript provides a comprehensive review of superconducting spintronics driven by electron symmetry filtering and interfacial SOC. It emphasizes the critical role of a crystalline MgO barrier in selectively transmitting specific electronic states between V(100) and Fe(100). The manuscript also highlights how interfacial SOC enables symmetry mixing, allowing for the interaction between ferromagnetic and superconducting orderings through MgO(100). This mutual interaction, mediated by interfacial SOC, facilitates the conversion of spin-singlet to spin-triplet Cooper pairs. The work provides key insights into designing SOC-based superconductor-ferromagnet hybrid structures for advanced superconducting spintronic functionalities.

(*) farkhad.aliev@uam.es

(\textsuperscript{\textdagger}) These authors contributed equally to the work.

\end{abstract}

%\pacs{71.70.Ej,72.70.+m,74.45.+c}
\maketitle

\section{Introduction}\label{section:Introduction}

Superconducting spintronics is an emerging field focused on exploring the hybridization of two \textit{antagonistic} order parameters: magnetism and superconductivity; and developing versatile cryogenic devices in which superconducting quantum coherence can be magnetically controlled.
%~\cite{Cai2022}.
To date, most progress in superconducting spintronics has relied on non-epitaxial superconductor/ferromagnet (S/F) hybrids, where natural or engineered magnetic textures have been used to convert singlet Cooper pairs into long-range spin-triplet states (LRTs) \cite{keizer-nature-06,Eschrig2011,Linder2015,Eschrig2015}. A few notable exceptions include epitaxially grown rare-earth-based twisted ferromagnetic layers, which offer controlled spin textures and spin-active interfaces~\cite{Robinson2010}, and perovskite-based S/F hybrids, where the high spin polarization of the ferromagnet originates from its electronic band structure~\cite{Santamaria2005}.

Although technically challenging, the growth and investigation of fully epitaxial S/F hybrids opens new avenues in superconducting spintronics, revealing physical mechanisms beyond those anticipated by existing theories. Most notably, the epitaxy allows to introduce a new versatile tool: electron spin and orbital symmetry filtering. Symmetry-dependent spin-polarized currents enabled by epitaxial growth have been central to both the prediction and discovery of giant tunnelling magnetoresistance (TMR) in Fe/MgO/Fe tunnel junctions~\cite{Butler2001,Mathon2001,Parkin2004,Yuasa2004}. Incorporating symmetry filtering into superconducting spintronics can can significantly strengthen the influence of interfacial spin-orbit coupling (SOC) between the S and F order parameters. In the case of Fe/MgO interfaces, the symmetry filtering mechanism is known to yield exceptionally high spin polarization, up to $87\%$~\cite{Lu2009}. Previously, symmetry filtering in S/F hybrids was only discussed in epitaxial Fe(100)/V(100) multilayer heterostructures in the context of the emergence of large magnetoresistance MR near the superconducting critical temperature ($T_C$)~\cite{Moodera2008}. Similar conclusions (although with smaller superconducting spin-valve effects) were reported later for V/Fe/V/Fe/CoO epitaxial multilayers~\cite{Nowak2012}.

On top of this, the perfect epitaxy has a number of crucial effects on the interaction between the F and S parts of hybrid devices.
First, it may enhance electron interference effects in transport. Coherent phenomena such as MacMillan and Tomasch resonances have been used to demonstrate long-range penetration of LRTs into ferromagnets~\cite{Visani2012,Costa2022}.
Second, it enables precise control over the magnetocrystalline anisotropy (MCA) of the F layer. This requires atomically sharp interfaces, only achievable through epitaxial growth, which also facilitates the fine-tuning of interfacial spin textures, including perpendicular magnetic anisotropy~\cite{Ikeda2010}. This in turn strongly influences the coupling between the S and F order parameters.
Third, epitaxial growth makes it feasible to insert thin insulating barriers (I) between the S and F layers, forming S/I/F structures. Though seemingly counterintuitive, these structures can host strong interfacial Rashba and Dresselhaus SOC. The reduced density of interfacial defects in these hybrids could help to distinguish the contributions of Rashba and Dresselhaus SOC \cite{Costa2021} and investigate shot nose. 
Last but not least, the epitaxy offers superior homogeneity and reproducibility, critical for meaningful comparisons with theoretical models (typically based on ideal, defect-free interfaces) as well as for reliable fabrication of devices such as Josephson junctions and S/F/S-based qubits, where interfacial SOC is expected to play a pivotal role~\cite{Cai2022}.

This manuscript presents a detailed examination and discussion of the current progress in normal and superconducting spintronics in epitaxial V/MgO/Fe-based hybrids, with a focus on the crucial role of electron symmetry filtering and interfacial SOC in these systems. The current introductory Section~\ref{section:Motivation} ends by motivating the work and with an exposition of the background that will be used during the rest of the review. Then, in Section~\ref{section:Growth} we provide a detailed explanation of the structure of the junctions under study and their fabrication process. Next, Section~\ref{section:RoomT} examines the normal-state magnetic and transport properties, including \textit{ab-initio} calculations of the interfacial SOC, the characterization of the junctions above $T_C$ and demonstrates how SOC already has a strong influence over the density of states (DOS) and transport. Section~\ref{section:LowT} discusses techniques to characterize the SOC-induced coupling between the ferromagnetic and superconducting states. Specifically, we discuss the spin-triplet superconductivity-induced change in the MCA, the observation of magnetoanisotropic low-bias conductance (MAAR), and superconducting quasiparticle-induced electron-electron interference in the ferromagnetic electrodes. Section~\ref{section:SN} summarizes some of the latest results in these type of junctions, which offer a distinct experimental evidence of LRTs, particularly through the observation of giant subgap shot noise. Finally, Section~\ref{section:JJs} offers an introduction into the current line of work: our first attempt to investigate electron transport in S/F/S junctions with \textit{lateral} geometry, which could provide more direct evidence for the elusive LRTs by enabling the observation of a magnetization-controlled Josephson effect.

\subsection*{Superconducting spintronics with SOC} \label{section:Motivation}

Spin-orbit coupling is an intra-atomic relativistic interaction between an electron’s spin and its orbital motion, which plays a key role in solid-state systems~\cite{Zutic2004}. While the importance of SOC in the generation of spin currents in spintronics is now well established~\cite{Jo2024}, its role in generating spin supercurrents in S/F hybrid structures was only recognized in the past decade~\cite{Bergeret2013,Banerjee2018,Martinez2020,Amundsen2024}. In such S/F systems, SOC enables a range of phenomena not present in homogeneous materials, including the emergence of Majorana zero modes~\cite{Manchon2015}, the paramagnetic Meissner effect~\cite{Espedal2016}, strong anisotropy in low-bias conductance (magneto anisotropic Andreev reflection MAAR)~\cite{Hogl2015,Vezin2020,Martinez2020}, and giant thermoelectric response~\cite{feng-nanoenergy-24,ouassou-prb-22,GonzalezRuano2023, Tuero2025}. It also influences $T_C$, which becomes dependent on the magnetization orientation~\cite{Buzdin1999,Banerjee2018}, and could give rise to non-reciprocal transport effects such as the superconducting diode behavior~\cite{spin-SDE,Ili2024}. Moreover, SOC facilitates the generation of LRTs, significantly affecting spin dynamics in S/F hybrids~\cite{Jeon2020}. SOC has also been predicted to enable magnetization-controlled Josephson effects in S/F/S junctions, where it mediates the interaction between the superconductor and the ferromagnet~\cite{Costa2017,eskiltprb2019,Bujnowski2019}. Additionally, SOC could modify the MCA in epitaxial ferromagnets~\cite{GonzalezRuano2020,GonzalezRuano2021}. Since the SOC strength can be electrically tuned, it offers promising avenues for the advancement of superconducting spintronic devices and realizing topological quantum computing~\cite{Cai2022}.

The main motivation behind the investigation of the V/MgO/Fe-based epitaxial hybrid structures, as reviewed here, arises from three key features that distinguish these S/F hybrids from practically all previously studied superconducting spintronic systems. First, the Fe/MgO interface and, to a lesser extent, the V/MgO interface induce a strong SOC (primarily Rashba, due to the breaking of the translational symmetry in the atomic lattice at the interface). Second, the Fe/MgO system provides a spin polarization up to $80\%$, closely approaching that of a fully spin-polarized ferromagnet in the Fe electrode. Third, as discussed earlier, the symmetry mismatch between the electronic structures of V(100) and Fe(100) creates a scenario in which virtually all electrons transferred between the ferromagnet and superconductor experience SOC, a unique situation in superconducting spintronics. Electrons whose symmetry is transformed from $\Delta_2$ to $\Delta_1$ via SOC are efficiently transmitted through the MgO barrier. 
%This symmetry selective transmission mechanism is similar to the well known underlying effect responsible for the record-high tunneling magnetoresistance (TMR) observed at room temperature in Fe/MgO/Fe magnetic tunnel junctions (MTJs)\cite{Yuasa2004,Parkin2004}.

\section{Sample growth and characterization}
\label{section:Growth}

\begin{figure}[b]
\begin{center}
\includegraphics[width=\linewidth]{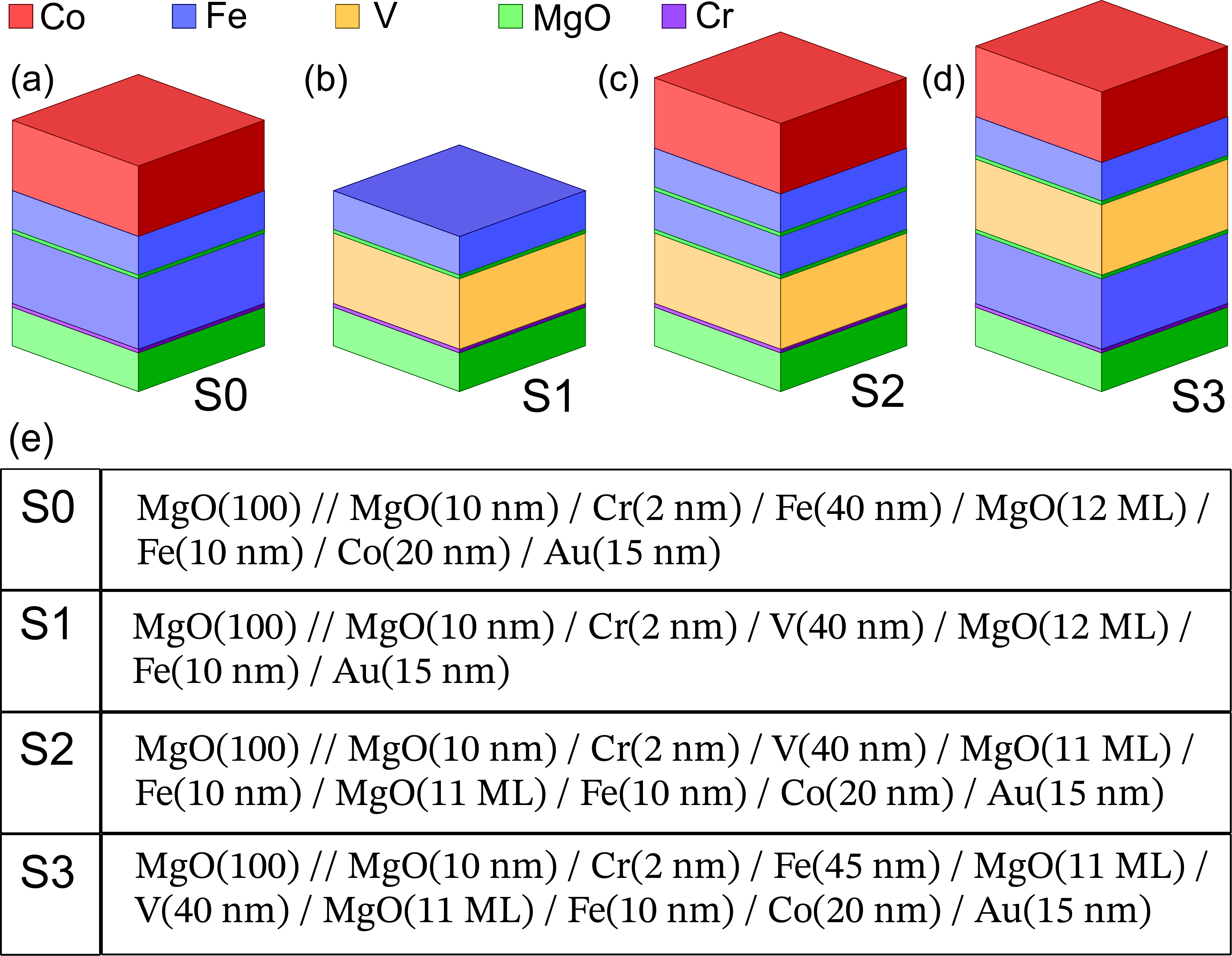}
\caption{Sketch of the structure of the types of junctions studied. (a) F/F, (b) N(S)/F, (c) N(S)/F/F, (d) F/N(S)/F. The different colors represent each material (see legend). The Au top layer is not shown as it is the same in all the samples. (e) Table with detailed information of the vertical structure of the samples.}
\label{fig:samples}
\end{center}
\end{figure}

Four sets of epitaxial single-crystalline samples have been fabricated using molecular beam epitaxy (MBE) in a chamber operating with a base pressure of approximately $1.1×\times10^{-10}$~Torr. Sketches of their vertical structures are illustrated in Figure~\ref{fig:samples}a-d, accompanied by the detailed multilayer structural sequence of each of the samples in Figure~\ref{fig:samples}e.
%\setkeys{Gin}{draft=false}
%\setkeys{Gin}{draft}
More details on the growth and characterization of the standard vertical MTJs (sample S0 which is Fe(40 nm)/MgO(12 monolayers)/Fe(10 nm)/Co(20 nm) could be found in Ref.\cite{Tiusan-2007}.

The first type of S/F hybrid samples (S1) consists of a V(40~nm)/MgO(12~ monolayers)/Fe(10~nm) MTJ, with vanadium as the bottom electrode and iron as the top electrode (Figure \ref{fig:samples}b). The second samples (S2) are a double tunnel junctions comprising a bottom V electrode and two ferromagnetic Fe electrodes (middle and top), separated by single-crystal MgO barriers (Figure \ref{fig:samples}c). The third samples (S3) are double-barrier MTJs with Fe as the bottom and top electrodes and V as the middle electrode (Figure \ref{fig:samples}d). In all three cases, the vanadium layers become superconducting at low temperatures ($\sim$~4~K). All samples were grown on (100)-oriented MgO substrates that were degassed at 650~°C for 30 minutes. During the subsequent growth of a 10~nm MgO seed layer, the substrate temperature was maintained at 450~°C. This seed layer serves to suppress residual carbon segregation at multilayer interfaces \cite{Tiusan-2007}.

\begin{figure}
\begin{center}
\includegraphics[width=\linewidth]{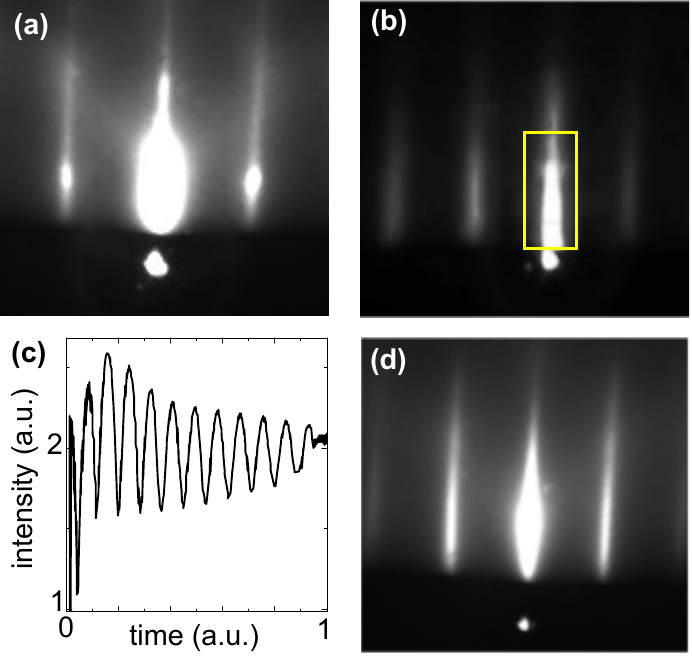}
\caption{Monitoring and characterization of the growth procces of the sample S1. (a)~Typical reflection high-energy electron diffraction (RHEED) image for the annealed V (40nm) layer. (b) RHEED image. The yellow square marks the area averaged for the (c) intensity oscillatons during the growth of the MgO barrier on top of V are shown. (d) RHEED image for the Fe layer on top of the MgO barrier. All images are taken from [110] azimuth of the MgO substrate.}
\label{fig:RHEED}
\end{center}
\end{figure}

For S1 and S2, a 2~nm Cr seed layer was deposited at $\sim$~30 °C to improve the wetting and the growth of the subsequent V(40~nm) electrode. This V layer was deposited at room temperature and subsequently annealed at 450~°C for 30 minutes to enhance crystallinity, as illustrated by the reflection high-energy electron diffraction (RHEED) images in Figure~\ref{fig:RHEED}a. The MgO tunnel barrier was then epitaxially grown at $\sim$~100 °C (Figure~\ref{fig:RHEED}b), with its thickness precisely monitored using RHEED oscillations (Figure~\ref{fig:RHEED}c). A~10 nm Fe electrode was deposited at the same temperature and annealed for crystallization and flattening at 400~°C for 20 minutes (Figure~\ref{fig:RHEED}d). For S1, the growth was completed with a 15~nm Au capping layer.

In the case of S2 (see Figure \ref{fig:samples}c), a second MgO barrier was grown epitaxially on the Fe (10~nm) electrode at $\sim$ 80 °C, followed by deposition of a second Fe (10~nm) electrode at a similar temperature. This layer was annealed at $\sim$~400~°C for 20 minutes. To magnetically harden the top electrode, a thin Co layer was deposited at $\sim$~100~°C, forming an epitaxial hexagonal structure with its c-axis tilted in the plane of the Fe layer. The Fe/Co bilayer enables the independent control of the two magnetic electrodes of the double-barrier junction under an external magnetic field, thus establishing a hard–soft magnetic configuration. The stack was finally capped with a 15~nm Au protective layer.

%\setkeys{Gin}{draft=false}
%\setkeys{Gin}{draft}

For S3 (Figure \ref{fig:samples}d), the multilayer growth started on an MgO(100)//MgO(10~nm) base template. A 45~nm Fe bottom electrode was deposited at $\sim$~30~°C and annealed at 500~°C for 30 minutes to improve the crystallinity and achieve an atomically flat surface. A first single crystal MgO barrier was then epitaxially grown layer by layer at $\sim$~70~°C, with its thickness monitored via RHEED oscillations. On top of this barrier, a 40~nm V electrode was epitaxially grown at the same temperature and annealed for flattening at 450~°C for 30 minutes. A second single crystal MgO barrier was subsequently epitaxially deposited at $\sim$~100~°C, followed by the growth of a 10~nm Fe electrode at $\sim$~90~°C. This electrode was annealed at 400~°C for 20 minutes to promote crystallization and surface smoothing. Similarly to S2, the top Fe electrode was magnetically hardened by the deposition of a 20~nm Co overlayer at the same temperature. The final structure was capped with a 15~nm Au layer for protection.

Following the MBE growth, all samples were patterned into in square-shape junctions of a size from $10\times10$ to $40\times40$~$\mu{\text{m}^2}$ (with the diagonal along the MgO substrate [100] direction) using UV lithography and ion etching, with the etching process being monitored step by step by in-situ Auger spectroscopy (see Ref. \cite{Tiusan-2007} for more details).

\section{Junction properties in the normal state}\label{section:RoomT}

We begin with a discussion of the junctions normal-state properties, including their non-volatile magnetization configurations and tunneling magnetoresistance, with special emphasis on the surprising TMR enhancement under applied bias observed in the S2 type junctions.
 
It is well known \cite{Hallal2013} that the interfacial SOC provides a perpendicular magnetic anisotropy (PMA) in MgO/Fe structures, responsible for the perpendicular magnetization alignment in thin (below 2~nm) films \cite{Ikeda2010}. For thicker Fe films, PMA is expected to compete with in-plane magnetic anisotropy. As a consequence, for the most part of this work, we selected a 10~nm thick magnetically free Fe layer, which enables three orthogonal and non-volatile magnetic states (two in-plane and one out-of-plane) without requiring an external magnetic field. As we show in the next chapters, these configurations allow us to systematically study the influence of the magnetization direction on the response of S/F hybrids (without the application of any external magnetic field), experiments that were previously unfeasible in superconducting spintronics.

\subsection{Non-volatile magnetization states}\label{section:MagneticStates}

In our junctions, the different non-volatile magnetic configurations are achieved by applying and removing an external magnetic field, using a 3D vector magnet composed of three superconducting coils. The three axes of the magnet are aligned with the main crystallographic directions of the junctions, and the applied field can reach 1~T in two of the axes and 3.5~T in the other one. 
In the N/F/F MTJs (S2 in Figure~\ref{fig:samples}), one of the electrodes is made to be a soft F (10~nm Fe) and the other a hard F (Fe/Co). The relative alignment of the two F layers is found by measuring the conductance at low biases (\textless10~mV). The magnetization's behavior (remanence, coercive fields, etc) depends on the competition between crystalline, volume (or shape) and surface anisotropies, as well as the interaction between both F layers. Due to the growth of the samples, the plane of the layers and the crystalline Fe bcc easy axis are the same (100)\cite{Tiusan-2007}. Because of that, in addition to the shape anisotropy, the coercive fields of the F layers are lower in the in-plane directions than in the out-of-plane one, and thereby the F magnetization will follow much more easily the external magnetic field if it is applied in-plane.

%\setkeys{Gin}{draft=false}
\begin{figure}
\begin{center}
\includegraphics[width=\linewidth]{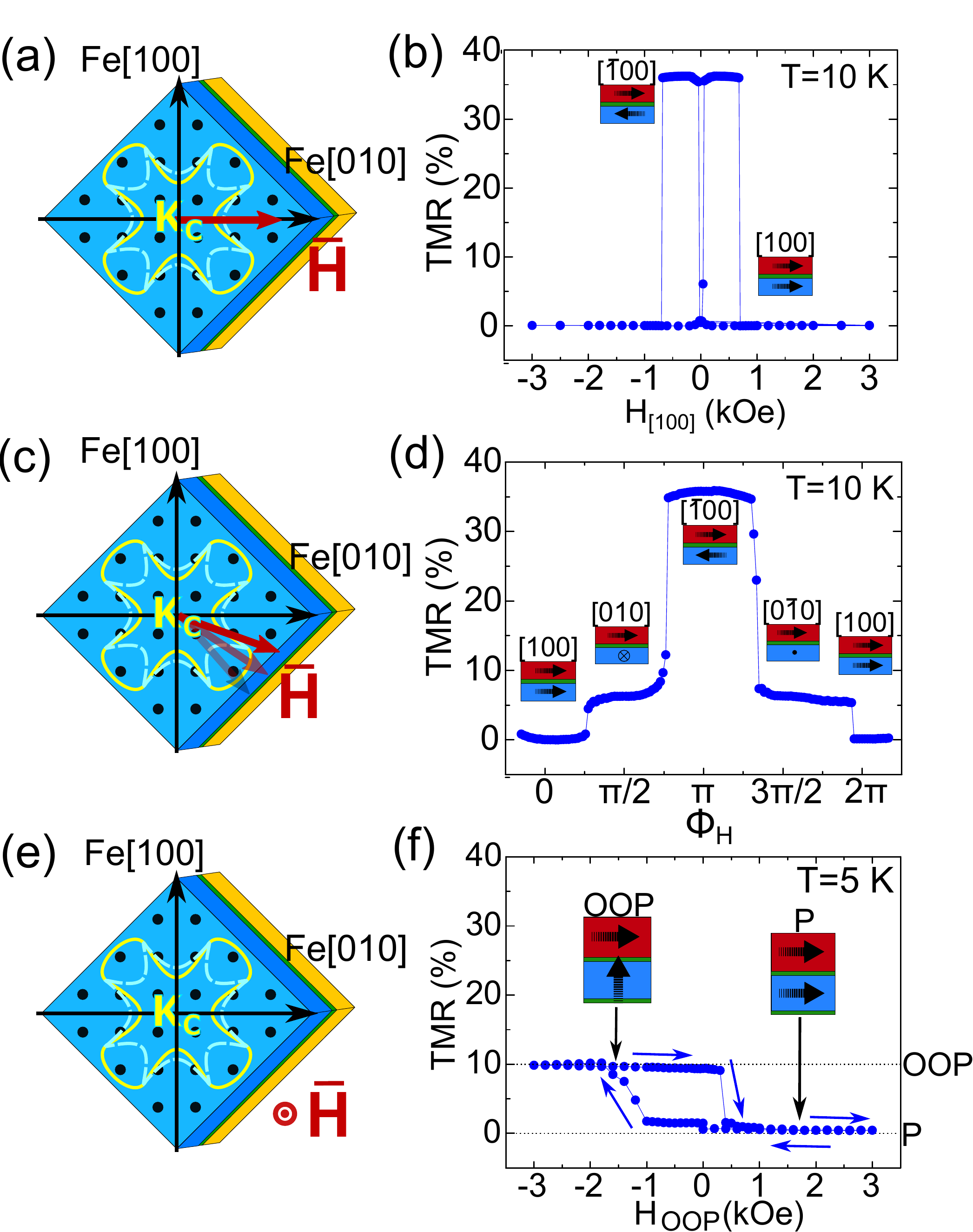}
\caption{(a,b) In-plane TMR in the easy axis [100] for a a N/F/F junction. Both parallel and anti-parallel magnetic alignments are observed, with a very low coercive field ($\sim25-50$~Oe) for the soft layer and a higher ($\sim400-600$~Oe) for the hard one. (c,d) In-plane rotation of the magnetic field, using a field higher than the soft layer coercive field (80~Oe), making it reorient following the field and magnetocrystalline anistropy (MCA), but lower than the hard layer coercive field, so it is fixed during the experiment. (e,f) Out-of-plane TMR at $T=5$~K with an out-of-plane alignment of the soft layer. Panels (a,c,e) sketch the orientation of the magnetic field with respect to the MCA in each of the experimetns. Figure adapted from refs.~\cite{GonzalezRuano2020,GonzalezRuano2021} with the necessary permissions.}
\label{fig:magnetic-states}
\end{center}
\end{figure}
%\setkeys{Gin}{draft}

We already mentioned that in order to study the interaction between S and F, it is important to have different remanent magnetic configurations available. Figure~\ref{fig:magnetic-states}a shows a TMR experiment in the N/F/F sample (S2) where the magnetic field ($H$) is swept along the Fe [100] crystallographic direction, where bias-dependent TMR is defined in terms of the relevant conductance, $G$ as function of an applied bias, $V$:
\begin{equation}
\label{TMR:equation}
\text{TMR}(V)=\dfrac{G_P(V)-G_{AP}(V)}{G_{AP}(V)}.
\end{equation}
It reveals the existence of a parallel (P) and antiparallel (AP) in-plane states of the two ferromagnetic electrodes. The coercive fields of the soft and hard magnetic layers can also be extracted, being 25-50~Oe and 400-600~Oe, respectively. The in-plane rotation sketched in Figure \ref{fig:magnetic-states}b shows the three possible in-plane magnetic configurations: the above mentioned P and AP states and the perpendicular in-plane (PIP) state. By making a rotation with an external field lower than the hard F coercive field, we maintained that magnetization fixed. The soft F magnetization follows the external field in jumps between four different directions which correspond to the direction of the Fe crystallographic easy axis. 
The lower effective TMR (around 40 $\%$) value is not due to the crystalline structure’s quality, but arises from the in-series configuration of the N/F/F junction which includes a normal metal electrode (vanadium), providing a nearly constant resistance which limits the overall TMR ratio.

Figure~\ref{fig:magnetic-states}c shows an experiment similar to that in panel~(a) but with $H$ applied perpendicularly to the plane of the sample. It can be deduced that there is an additional out-of-plane (OOP) magnetization state in which the hard layer remains magnetized IP and the soft layer is OOP. The transition from the P to the OOP state occurs at $H_{OOP}\simeq1.5$~kOe. Importantly, this OOP state is remanent: it is persistent when $H_{OOP}$ is removed, and a magnetic field of 400~Oe in the opposite direction is needed to switch back to the P in-plane state.
%In the absence of metastable states, the equilibrium orientation of the magnetization of a thin F film is a result of the competition between the crystalline, shape (or volume) and surface anisotropies. The crystallographic axis in our junctions matches the plane of the layers, although its influence on the magnetization orientation is not so important for the out-of-plane direction as the competition between shape and surface anisotropies. If the thickness of the F layer substantially exceeds values of about 10~nm, the shape anisotropy dominates and tends to align the magnetization in-plane. In contrast, if the film is thinner than 2~nm the surface anisotropy tends to align the magnetization out of plane, leading to out of plane magnetization alignment. 
Recent trends in spintronics using MTJs take advantage of PMA to provide large perpendicular room temperature TMR, enhanced thermal stability\cite{Ikeda2010}, low spin-torque switching currents\cite{Leutenantsmeyer2015,Lau2016}, and record small lateral sizes~\cite{Igarashi2017}. As we have shown above, our 10~nm thick magnetically free Fe layers (due to competing crystalline, shape and surface anisotropies) enable us to change the magnetization of the film either IP or OOP so that the magnetization stays remanent at low temperatures when switching off the external field, as it is shown in Figure~\ref{fig:magnetic-states}c.

To elucidate the mechanism behind the different magnetization orientations due to competing anisotropies we carried out micromagnetic simulations MuMax3\cite{mumax}, considering three models that differ in their in-depth anisotropy. The simulations described in Ref.\cite{Martinez2018} show that the spin-flip transition is rather weakly affected by the presence of the Fe layer cubic anisotropy. This demonstrates that it is mainly a demagnetization energy competing with PMA which provides the spin reorientation transition.

\subsection{\textit{Ab-initio} SOC calculations at the V/MgO and Fe/MgO interfaces}\label{section:SOCCalculation}

\begin{figure}
\begin{center}
\includegraphics[width=\linewidth]{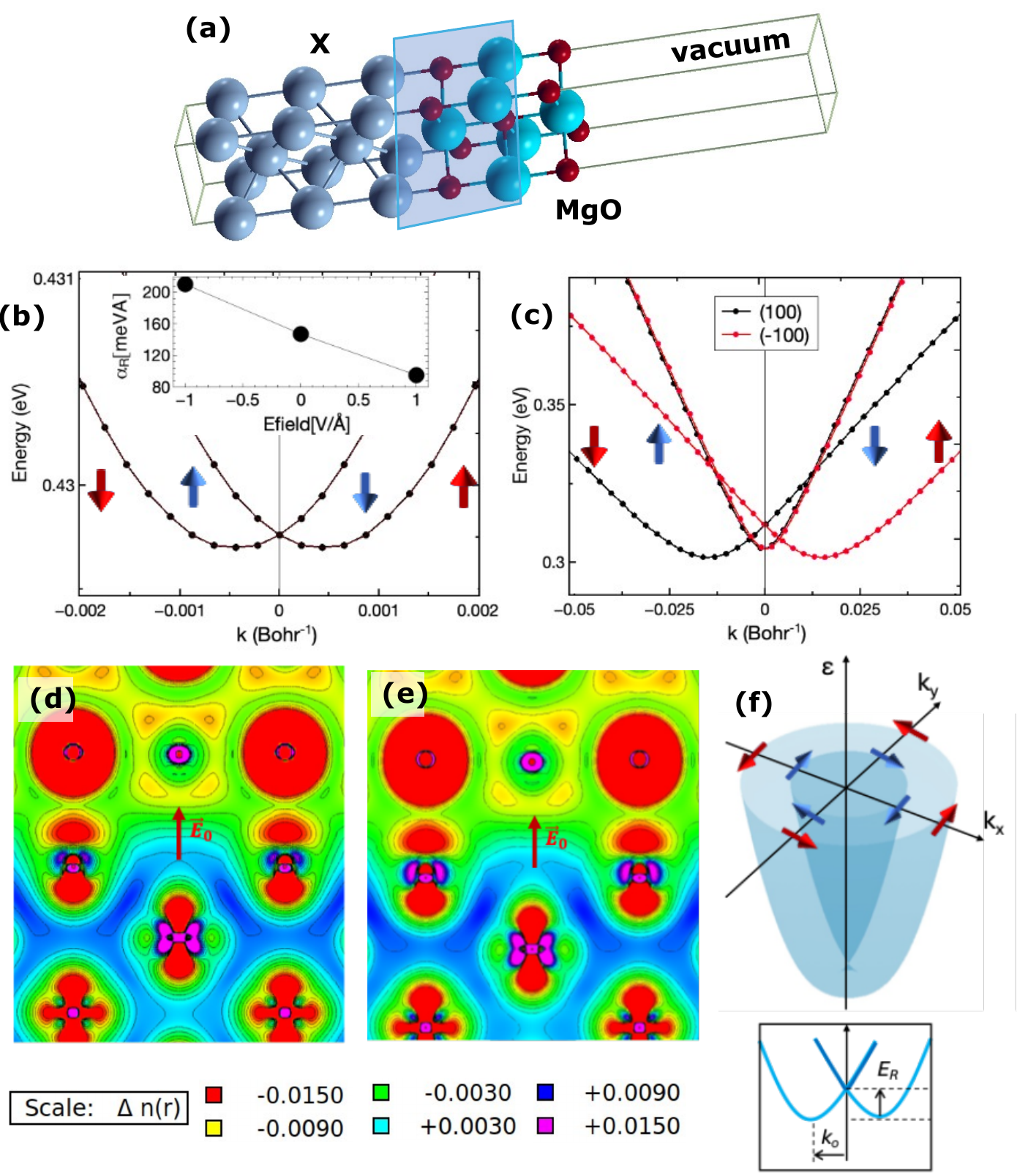}
\caption{(a)~Supercell model for the X/MgO interfaces with X = V, Fe. Calculated band splitting due to Rashba spin-orbit coupling at the (b) V/MgO and (c) Fe/MgO interface in the absence of electric field. Inset of (b): variation of $\alpha_R$ with the electric field in case of V/MgO interface. Charge density at the X/MgO interface (blue rectangle in (a)) for (d) V/MgO and (e) Fe/MgO. (f) Sketch model for the Rashba splitting of the parabolic bands, the $k$ lines in (b) and (c) correspond to $k_y=0$, the direction in the k space (100) corresponds to $k_x>0$ and (-100) to $k_x<0$.}
\label{fig:AbInitio}
\end{center}
\end{figure}

To further understand the underlying electronic structure of our system, the theoretical study of the Rashba SOC and its electric field dependence has been performed using the \textit{ab-initio} full potential linear sugmented plane wave FP-LAPW code WIEN2k~\cite{Wien2k} within a fully relativistic spin-orbit scheme. In our calculations, we used a supercell model thoroughly chosen to describe the X/MgO(001) interface, where X = V and Fe (see Figure~\ref{fig:AbInitio}a). The Brillouin zone was sampled using a $25\times25\times1$ $k$-point mesh, which was selected based on a preliminary convergence study of the total energy with respect to the number of $k$-points. Given the strong sensitivity of such calculations to the energy convergence parameters in WIEN2k, particularly when relativistic local orbitals (LAPW+lo) are included, we employed $E_{max}=100$~Ry where $E_{max}$ is the maximum energy cutoff for plane waves, controlling the size and accuracy of the basis set. This choice ensures that all scalar-relativistic eigenstates are taken into account when SOC is enabled. The basis set size for the wave function expansion was chosen as $R{\times}K_{max}=7$, where $R$ is the smallest muffin-tin radius, $RMT$, and $K_{max}$ is the largest reciprocal lattice vector. Following the zig-zag potential electric field implementation in the WIEN2k code introduced by Stahn et al.~\cite{Stahn2001}, the supercell is constructed such that one half contains the multilayer sequence, while the other half is a vacuum region. This setup generates a constant electric field, $E=-\Delta{V}/\Delta{z}$, localized at the X/MgO interface. Both the amplitude and the sign of the field are determined by the applied potential ramp and whether the ramp is ascending or descending within that region. Therefore, for different values of the electric field, we calculated the Rashba parameter $\alpha_R$ from the band structure obtained from WIEN2k.

%\setkeys{Gin}{draft=false}
%\setkeys{Gin}{draft}

The spin-orbit term of the non-relativistic Dirac (Pauli) Hamiltonian:
$H_{SO}=\sfrac{\hbar}{(2m_0c)^2}\overrightarrow{\nabla} V\cdot(\overrightarrow{\sigma}\times\overrightarrow{p})$ can be implified for 2 dimensional electronic systems with the confinement direction (e.g., $\overrightarrow{OZ}$ in a Cartesian system) perpendicular to the propagation direction (within the $XOY$ plane) leading to the Rashba Hamiltonian: $H_R=\alpha_R\overrightarrow{\sigma}\cdot(\overrightarrow{k}\times\overrightarrow{e_z})$ where $\alpha_R=\sfrac{\hbar^2}{(2m_oc)^2}{\partial{V}}/{\partial{z}}$ is the Rashba constant which is a measure of the SOC strength and $\overrightarrow{e_z}$ is the unit vector of the $\overrightarrow{OZ}$ (electron confinement) direction. From its definition, $\alpha_R$ is proportional to the electric field $E=-{\partial{V}}/{\partial{z}}$ aligned along the $\overrightarrow{OZ}$ direction. At the interface between two dissimilar materials, interfacial electric fields naturally arise, giving rise to Rashba-type SOC effects. The strength of the SOC is further enhanced at metal surfaces, where the breaking of translational symmetry is equivalent to a potential gradient experienced by the electrons. In multilayer heterostructures comprising a metal–insulator or metal–semiconductor interface, the formation of a depletion region further induces a pronounced interfacial electric field. By diagonalizing the Rashba Hamiltonian, the eigenvalues will be: $E_{\pm}(k_{ \parallel})=\sfrac{\hbar^2k_{\parallel}^2}{2m_0}\pm\alpha_R\vert{k}_{\parallel}\vert$ representing parabolic bands with an offset of the parabola minimum in positive (or negative) $k$ values (Rashba splitting). The minimum of the parabola can be found by ${\partial{E}}/{\partial{k}}=0\Rightarrow{k}_0={m_0\alpha_R}/{\hbar^2}$ so that $E_{min}=E_R=\sfrac{\hbar^2k_0^2}{2m_0}$ that leads to $E_R={k_0\alpha_R}/{2}$. From this equation, $\alpha_R$ can be calculated from the values of $E$ and $k$ corresponding to the minimum of the parabolic dispersion band: $\alpha_R={2E_R}/{k_0}$. We used this equation to extract the Rashba parameter from the Rashba offset of the parabolic bands calculated for the X/MgO/vacuum supercell models. Note that, experimentally, the SOC constant $\alpha_R$ can be extracted from angular `resolved photoemission (ARPES)~\cite{Nicolay2001, Reinert2001}.

In Figure~\ref{fig:AbInitio}b,c we illustrate the Rashba splitting of the parabolic bands for V/MgO and Fe/MgO interfaces, respectively, corresponding to zero electric field. From these bands, we extract the corresponding $\alpha_R$ parameters of $\alpha_R=147~\text{meV}$\si{\angstrom} (V/MgO) and $\alpha_R=730~\text{meV}$\si{\angstrom} (Fe/MgO). From Figure~\ref{fig:AbInitio}d,e illustrating the calculated charge density in the interface region, we clearly notice that the Rashba SOC is correlated to the interfacial charge depletion at the X/MgO interface and the corresponding intrinsic field $\overrightarrow{E_0}$. This depletion will be reduced (increased) in the presence of a positive (negative) electric field (see inset in Figure~\ref{fig:AbInitio}b corresponding to the V/MgO system). However, a very large electric field of about $1~\text{V}/$\si{\angstrom} $=10~\text{V}/\text{nm}$ is needed to have a significant effect on the intrinsic Rashba field. Within the range of applied bias voltages used in our experiments, no significant variation of the parameter $\alpha_R$ induced by the bias of the MTJ is expected. Therefore, to first order, the SOC is assumed constant across all investigated biases in the magneto-transport measurements. Only the effects of the electronic structure on conductance need to be considered, which can be adequately described within a rigid band model. In this framework, SOC mediates the mixing between the surface $\Delta_1$ and bulk $\Delta_2$ orbital symmetries in vanadium, thereby enabling electronic transport in the V/MgO/Fe system~\cite{gonzalez-ruano2022boost}. 

\subsection{SOC-induced conductance bottleneck}\label{section:bottleneck}

The orbital symmetries in the electronic structures of V(100) and Fe(100) are mismatched, implying that the heterojunctions based on these materials 
should exhibit negligible conduction under low-bias conditions. Indeed, early studies on fully epitaxial \textit{lateral} bcc Fe/V/Fe spin valves have already highlighted the critical role of orbital symmetries in spin-dependent transport across a V/Fe interface~\cite{Moodera2008}.

\begin{figure}
\begin{center}
\includegraphics[width=\linewidth]{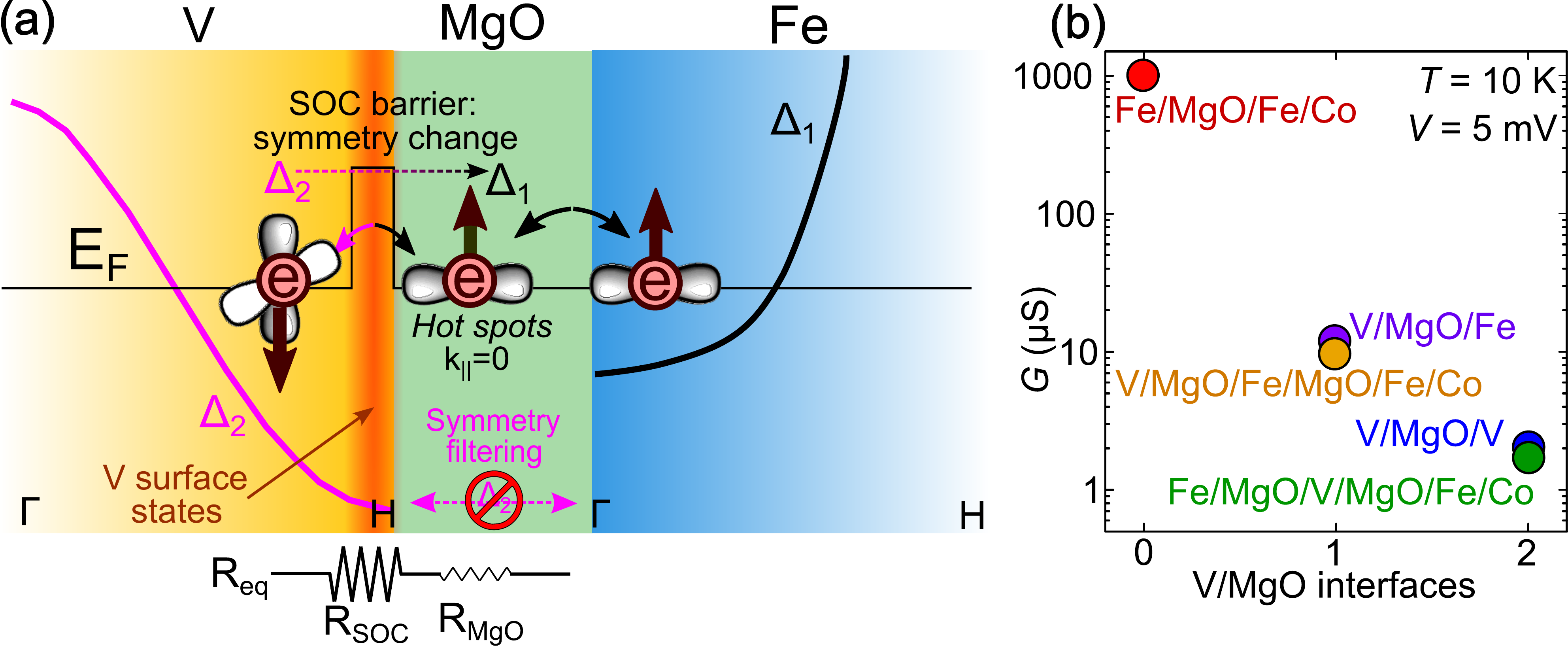}
\caption{(a)~Transport mechanism in the V/MgO/Fe barrier, sketching the vanadium mixed $\Delta_2$ and $\Delta_1$ surface states, interfacial spin-orbit coupling and symmetry filtering in the MgO. (b)~average conductance for different types of samples of the same lateral size ($20\times20~\mu\text{m}^2$), as a function of the number of V/MgO interfaces for each type.}
\label{fig:bottleneck}
\end{center}
\end{figure}

At the Fermi energy $E_F$, (100)V possesses $\Delta_2$ orbital symmetry, whereas Fe(100) is dominated by $\Delta_1$ symmetry (see Figure~\ref{fig:bottleneck}a). Therefore, MgO acts as an insulating layer that suppresses the $\Delta_2$ states and enables TMR in Fe(100)/MgO(100)/Fe(100) structures~\cite{Parkin2004,Yuasa2004,Butler2001,Mathon2001}. Nevertheless, the structural inversion asymmetry in our junctions introduces an interfacial SOC at the V/MgO interface, which works in conjunction with the effective $\Delta_2$ filtering barrier of crystalline MgO. As sketched in  Figure~\ref{fig:bottleneck}a and discussed in Ref. \cite{GonzalezRuano2021-HB-TMR} , SOC-driven spin-flip scattering mixes the $\Delta_2$ and the $\Delta_1$ states, opening a pathway for electron tunneling above $T_C$ at low bias, as depicted in Figure~\ref{fig:bottleneck}a.

Another channel contributing (in sequence) to the low bias transport in these epitaxial junctions originates from the so-called ``hot spots'' in the reciprocal space provided by the $\Delta_1$ states of electrons that undergo normal incidence to the barrier, allowing a strong transmission through MgO~\cite{Butler2001,Mathon2001}. The normal-state transport in this system can therefore be approximated by an effective resistance $R_{\text{eq}}=R_{\text{SOC}}+R_{\text{MgO}}$, where the resistance associated with the symmetry-enforced SOC barrier, $R_{\text{SOC}}$, is significantly higher than that the one arising from the conventional MgO barrier, $R_{\text{MgO}}$.

This framework, in which tunneling is governed by symmetry-based spin filtering rather than the barrier height, is supported by our measurements of the low-bias differential conductance $G$ (Figure~\ref{fig:bottleneck}b). Compared with junctions lacking such an interface, $G$ is reduced by roughly two (three) orders of magnitude in the presence of one (two) interfaces. This highlights a unique situation in (superconducting) spintronics in which practically all electrons that are transmitted between V and Fe undergo the effects of SOC.

%\setkeys{Gin}{draft=false}
%\setkeys{Gin}{draft}

\subsection{TMR enhancement with applied bias}\label{section:HighBiasTMR}

As we have already mentioned above, the orbital symmetry-controlled TMR was first predicted and then experimentally confirmed about two decades ago\cite{Butler2001,Mathon2001}. Today, it is probably the most notable phenomenon in the whole spintronics field, finding applications from magnetoresistive random access memories, logic gates\cite{Ney2003} and magnetic sensors \cite{Lenz2006}. TMR was first prediced to reach values exceeding $1000\%$ at $V=0$~\cite{Butler2001,Mathon2001}. Unfortunately, this is not possible in experimental setups due to structural defects and other effects, such as diffusion or roughness at atomic interfaces or emerging electronic states near the surfaces, which only allow for a $\sim200-600\%$ TMR in realistic cases\cite{Ikeda2008}.

%\setkeys{Gin}{draft=false}
\begin{figure}
\begin{center}
\includegraphics[width=\linewidth]{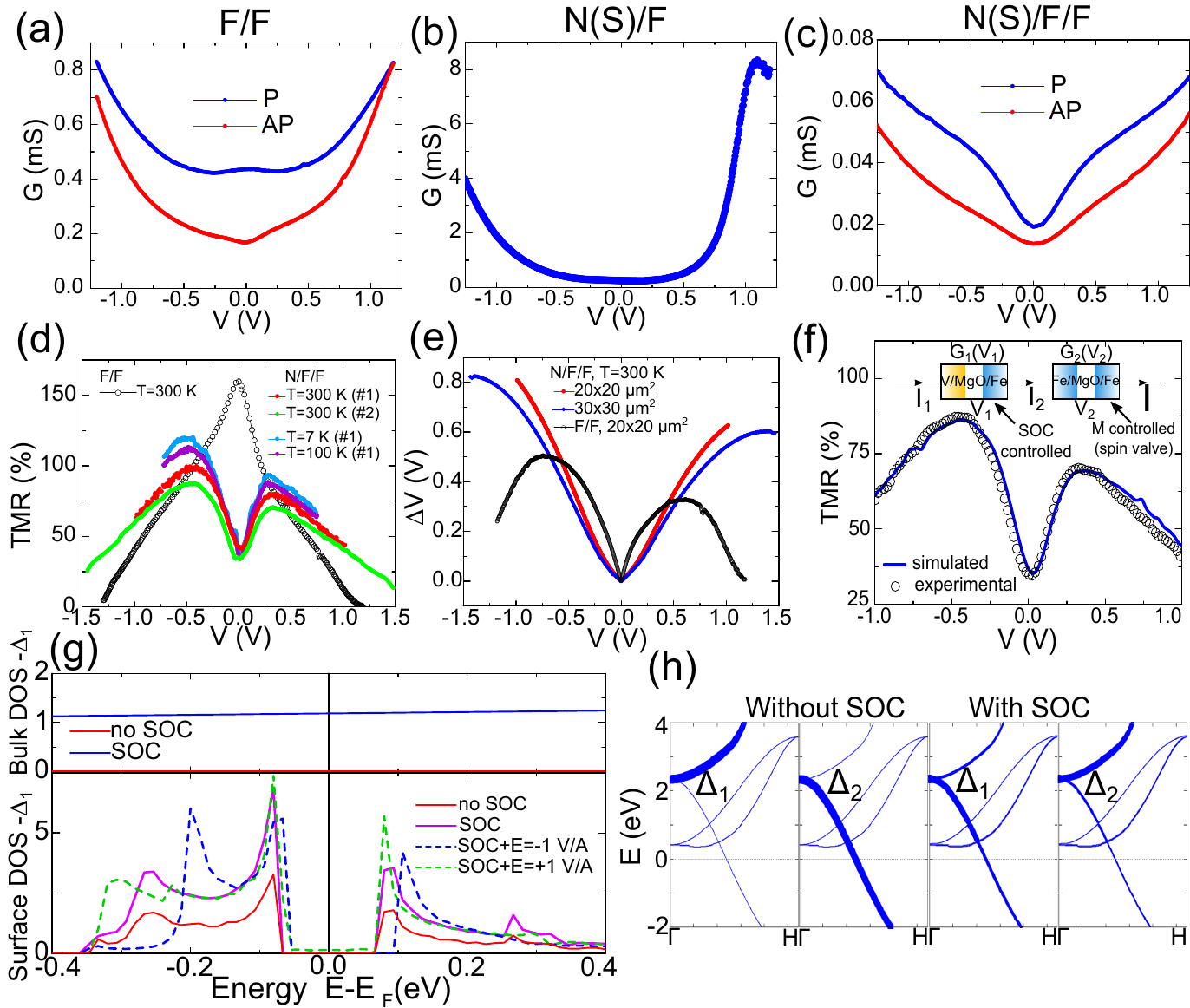}
\caption{(a), (b) and (c) show the conductances of the F/F, N/F and N/F/F junctions under study. (d) TMR vs applied bias for different junctions and temperatures. The blue, purple, red and green data are taken from two N/F/F junctions (legend \#1 and \#2). The black open dots data correspond to a F/F junction at room T. (e) $\Delta V$ vs applied bias in a F/F junction at room T (black), and for two different N/F/F junctions. (f) TMR($V$) of an example (typical) N/F/F junction, with the fitting of the model of two in-series resistors (see inset sketch). (g) $\Delta_1$ DOS in the bulk (upper part) and surface (below) parts of vanadium. Near the surface, the new DOS peak can be seen close to $\pm0.1$~eV from the Fermi level when SOC is present. (h) vanadium band structure, without SOC (first 2 panels) and with SOC (3 and 4). Here, each line's thickness is proportional to the indicated electronic symmetry. This was used to calculate the DOS in (g). Figure adapted from\cite{GonzalezRuano2021-HB-TMR} with the necessary permissions.}
\label{conductances-TMR-DV}
\end{center}
\end{figure}
%\setkeys{Gin}{draft}

Another major challenge that TMR has to overcome in MTJs is its degradation with applied bias \cite{Parkin2004,Yuasa2004,Waldron2006,Guerrero2007,Herranz2010}. In non-epitaxial devices where the electron tunneling is incoherent, this is due to electrons with extra energy above $E_F$, (so-called ``hot electrons" \cite{Parkin1997}), that are usually produced when bias or temperature are increased. Then, these  hot electrons produce magnon excitations near the interface where the tunneling takes place, which in turn result in spin-flip events during tunneling~\cite{Moodera1995}. The parallel configuration of the F electrodes (in standard F/F junctions) is not greatly affected by this since the spin-flips do not result in extra electrons tunneling, but the same effect has a much bigger influence on the bias-dependent conductance in the AP configuration, as here spin-flips allow the electrons to pass through the barrier, so the TMR (which measures the relation of the P and AP configuration conductances) is also reduced.
In epitaxial junctions, the coherent transmission between electronic bands is able to conserve the orbital symmetries in the tunneling process\cite{Guerrero2007}. When a bias is applied to the MTJ, this band-to-band transport has a stronger effect for the AP configuration than for the P one, reducing the TMR\cite{Waldron2006}.

Interestingly, the TMR of N/F/F (V/MgO/Fe/MgO/FeCo) junctions shows an unusual behavior under applied bias\cite{GonzalezRuano2021-HB-TMR}: it steadily increases from below 50\% to above 100\% with an applied bias below 0.5~V in a broad temperature range, from cryogenic to room temperature (Figure~\ref{conductances-TMR-DV}d). We have found that SOC was the main factor behind this effect, as it controls the tunneling of the N/F (V/MgO/Fe) part of the junctions (this same mechanism will still take place, accompanied by other processes, when we discuss transport in the superconducting regime in the following sections) while the F/F part is more standard, depending on the magnetic orientation of the two layers. In Figure~\ref{conductances-TMR-DV}d, we show a comparison of the TMR vs bias in a F/F (a) and a N/F/F (c) junction. As expected, TMR is higher at zero bias and reduces with applied bias in the F/F junction, while it is enhanced with applied bias in the N/F/F one, at least up to $V=0.5$~V.

When it comes to potential applications, the TMR is not the only figure of merit that is often examined. The output voltage, defined as

\begin{equation}
\Delta V=\left|V\right|\times\frac{R_{AP}-R_P}{R_P},
\end{equation}

is also an important factor. In the equation above, $R_P$ ($R_{AP}$) is the value of the electrical resistance in the P (AP) magnetic configurations. In our junctions, it improves the values of F/F junctions with similar structure at biases above $V=0.5$~V, as we demonstrate in Figure~\ref{conductances-TMR-DV}e. It is worth noting that, to the best of our knowledge, its value at room temperature (above $\Delta V=0.8$~V) sets a record in spintronics\cite{Tiusan2006}. At low applied biases, $\Delta V$ is even larger in F/F junctions, but it rapidly decreases with applied bias, in contrast to the steady increase in N/F/F junctions.

To find the explanation for this unusual behavior, we can first look at Figure~\ref{conductances-TMR-DV}b and notice a strong conductance increase that takes place with applied bias in the N/F junctions. Due to the aforementioned symmetry filtering, these MTJs are rather resistive at low biases. To give a comparison without the MgO-induced conductance bottleneck, control samples with an Au layer instead of a Fe one have a thousand times larger conductance when the lateral size is the same (see Ref.\cite{GonzalezRuano2025}, Supplemental Material). For the N/F/F junctions, we show $G$ in both the P and AP configurations of the F/F part in Figure~\ref{conductances-TMR-DV}c. Here one can observe how there is a range of biases where both conductances diverge. In Figure~\ref{conductances-TMR-DV}g, we sketched a region of surface states in the vanadium layer, which as we mentioned present electronic symmetries different than the rest of the layer.

These states were found with \textit{ab-initio} calculations similar to the ones described in Section~\ref{section:SOCCalculation}, and are crucial for understanding the TMR vs bias behavior, as these states allow us to explain the high zero bias resistivity of V/MgO/Fe: in vanadium, there is only one $\Delta_2$ electronic band crossing the Fermi level in the normal incidence direction. However, this symmetry does not exist at the Fermi level on the other side of the barrier, in the Fe electrode (if it would exist, the MgO barrier would stop the transmission\cite{Butler2001}). The low bias transport takes place only when $\Delta_2$ electrons change their orbital symmetry to $\Delta_1$ in the region with these surface states. A similar phenomenon was already published, where Fe/V/Fe/MgO/Fe junctions had much higher TMR than similar ones without the V layer\cite{Feng2009}.

The density functional theory (DFT) calculations are similar to the ones explained in the previous section. What the new calculations revealed was a peak in the vanadium surface density of states near the MgO, just below $V=100$~mV, which we found to agree with previous scanning tunneling spectroscopy experiments\cite{Bischoff2001}. The crucial part is that the calculations showed that these surface DOS would be enhanced if we added interfacial SOC. The reason for this is that, without SOC, the Fermi level falls right inside a gap for $\Delta_1$ states, which is no longer present in with SOC due to scattering processes allowing $\Delta_1$ and $\Delta_{2}$ to mix, as sketched in Figure \ref{fig:bottleneck}a.

With these calculation results, we are able to model the system with a relatively simple approximation that satisfactorily reproduced the TMR vs bias experimental results. The final result is depicted in Figure~\ref{conductances-TMR-DV}f, where we can observe a nice fit to the experimental TMR($V$) curve of a N/F/F junction. Specifically, we solve \cite{GonzalezRuano2021-HB-TMR} the nonlinear circuit equations by using Kirchhoff voltage and current laws (KVL, KCL): The KCL or charge continuity condition, $I_1=I_2=I$, allows us to determine the voltage drops $V_1$ and $V_{2(P/AP)}$ at each barrier from the individual $i(V)$ of each of them (measured experimentally). The KVL provides the total voltage drop on the serial device, $V=V_1+V_{2(P/AP)}$. The total conductance will be $G_{P/AP}(V)=I/V$, and then the corresponding TMR($V$) from Eq.\ref{TMR:equation} can be calculated. We also used a parametrized $G_1(V)$ curve for the N/F part, adjusted to obtain the best fit of the final TMR($V$) curve for a real N/F/F junction. The result, shown in Figure~\ref{conductances-TMR-DV}f , accurately matches the TMR vs. bias in the N/F/F MTJs.

\section{SOC-induced effects in S/F systems: low temperature experiments} \label{section:LowT}

This section explains how the generation of spin-triplet Cooper pairs in epitaxial S/SOC/F systems with spin and orbital symmetry filtering has been experimentally probed. Some effects discussed below have been anticipated theoretically. These predictions include: (i) magnetoanisotropic Andreev reflection~\cite{Hogl2015} and (ii) superconductivity-induced changes in MCA\cite{Johnsen2019}. As further indirect evidence, we also use above-gap conductance anomalies (MacMillan resonances~\cite{MRR1}) to verify the possible contributions of SOC and magnetic spin textures to the generation of LRTs\cite{Visani2012}.

\subsection{Magneto-anisotropic Andreev reflection}\label{section:MAAR}

In this section, we will see how the low bias MR in all-epitaxial S/F MTJs reveals a substantial modification below the superconducting transition, gaining a three orders of magnitude higher MR anisotropy. The non-volatile magnetic configurations (IP and OOP) allow us to rule out orbital and vortex effects and identify the SOC origin of the observed MR. Such MR reaches $\sim20\%$ without an applied magnetic field, and is further increased for large magnetic fields~\cite{Martinez2020} .

In MTJs with SOC, one only needs one ferromagnetic layer to achieve MR. This is the so-called tunneling anisotropic MR (TAMR)~\cite{Fabian2007}. Unfortunately, its usual values are of around $\lt1\%$~\cite{Lu2012}, which for most practical purposes is useless. In the S/F MTJs under study, we found a huge increase in the MR of the Fe layer when the V is cooled down below $T_C$, entering the superconducting state. As we will see, interfacial SOC plays a central role, suggesting the presence of magnetization-dependent LRTs formation. Notably, the materials that these MTJs are composed of are readily available and well-tested in commercial uses, since they are based on Fe/MgO junctions~\cite{Parkin2004,Yuasa2004,Cascales2012}.

To observe these effects, one must first identify a distinct feature of interfacial SOC: above $T_C$, the influence of interfacial SOC has observable effects on the TAMR of N/F junctions, resulting in magnetization-dependent conductance~\cite{Fabian2007}. This effect by itself is too faint for any potential uses (usually $\text{TAMR}=[G(0)-G(\pi/2)]/G(\pi/2)\approx0.01\%$)\cite{Martinez2020}. However, exploring the conductance in the same junction under $T_C$ reveals a much higher anisotropy. In the superconducting regime, the counterpart for TAMR is MAAR~\cite{Hogl2015}. Both have the same definition, but for MAAR the conductance is considered for biases within the superconducting gap:

\begin{equation}
\text{MAAR}=\dfrac{G(0)-G(\theta)}{G(\theta)},
\end{equation}

\begin{figure}
\begin{center}
\includegraphics[width=\linewidth]{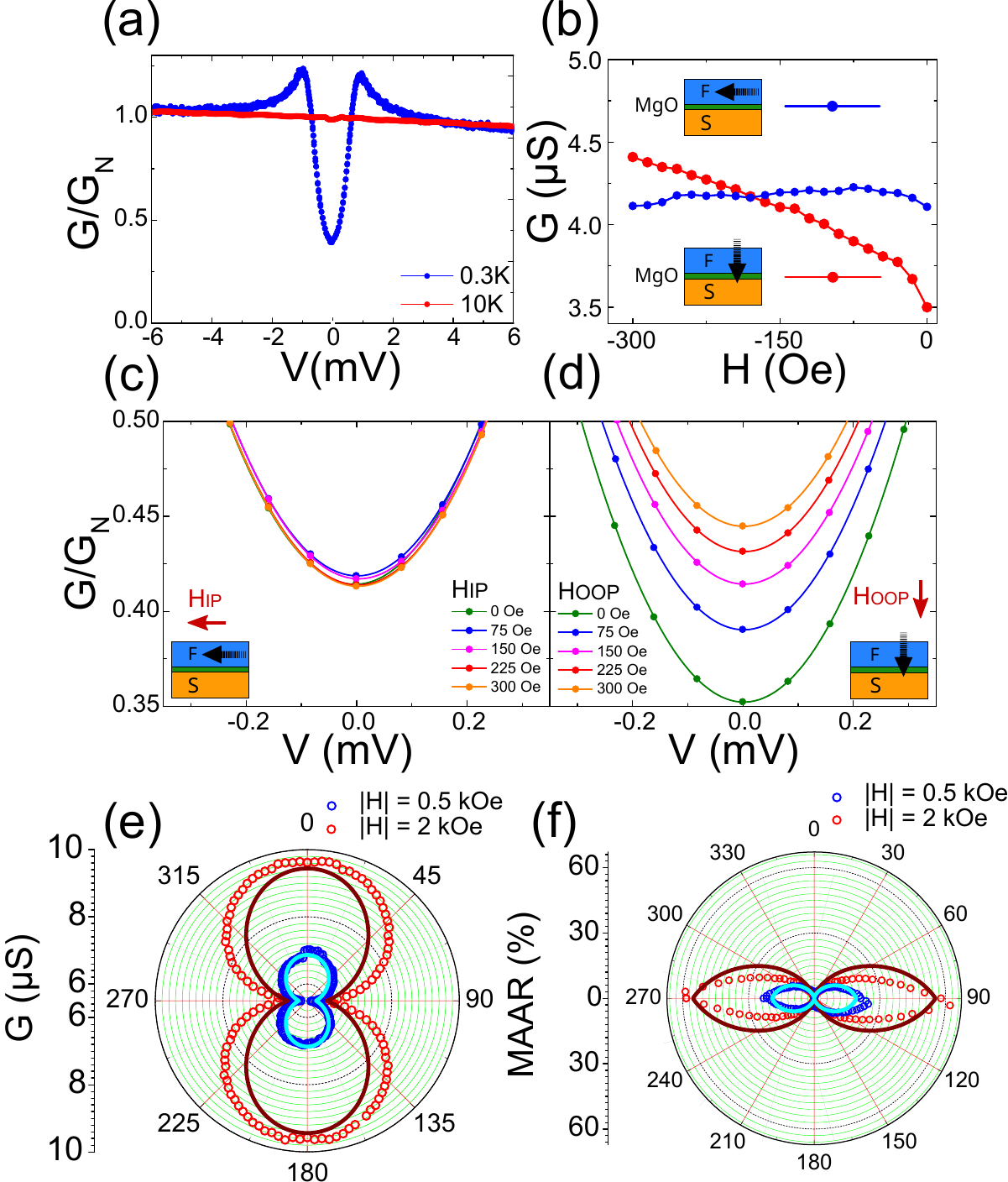}
\caption{(a) Conductance for V/MgO/Fe junctions at $H=0$, below (blue) and above (red) $T_C$. (b) Conductance at $V=0$ and $T=0.3$~K, vs. in-plane (blue) and out-of-plane (red) applied fields, showing a $\sim17\%$ anisotropy in the conductance at zero field (MAAR). (c) and (d) zoom into the conductance curves near $V=0$ that give the results in (b), where the conductance is normalized to the normal-state value at $V=3$~mV, G$_\text{N}$. (e) Anisotropy of the conductance at $V=0$ and $T=0.3$~K for out of pane field rotation  with module $H=0.5$~kOe (blue dots) and $H=2$~kOe (red dots), and the model overlaid (solid lines).  (f) MAAR extracted from (e). Here, experimental points are again shown as dots, while the model is depicted as solid lines. The rotation field spans orientations from in-plane (0º and 180º) to out-of-plane (90º, 270º) configurations. Figure adapted from ref.~\cite{Martinez2020} with permissions.}
\label{fig:MAAR-vs-H}
\end{center}
\end{figure}

Apart from the MAAR values, we can extract more information from the experimental data~\cite{Martinez2020}. Perhaps one of the most important parts is the transparency of the tunnel barrier. To model the experimental data, we used a BTK model~\cite{Hogl2015,Blonder1982} where the transparency of the barrier in the normal state is given by $1/(1+Z^2)$ (here, $Z=0$ means perfect transparency, while $Z\gg1$ is the tunnel junction limit). Therefore, the stronger (less transparent) the barrier is, the more the conductance will decrease~\cite{Blonder1982}. In this regard, Figure~\ref{fig:MAAR-vs-H}a shows a comparison of (normalized) conductances of the same sample below and above $T_C$ ($T=0.3$~K and $T=10$~K). From the curves, the conductance change can be used to extract a transparency which turns out to fall in an intermediate regime (Ref\cite{Martinez2020} discusses this in more detail), in between typical values among those of point contacts~\cite{Soulen1998,Nadgorny2012} and spin-polarized tunneling~\cite{Zutic2004,Parker2002,Parkin2004}.

Andreev reflections are also highly susceptible to SOC, in particular for high spin polarizations~\cite{Zutic2004} such as in our system. This is because, in the limit of full spin polarization, conventional Andreev reflection is not allowed, so the conductance drops to zero. This totally changes when interfacial SOC appears, as it allows for an analogue process called \emph{spin-flip Andreev reflection}~\cite{uti1999}. In the case of the V/MgO/Fe junctions, the Fe/MgO barrier has a very high effective spin polarization thanks to the symmetry filtering~\cite{Zutic2004,Parkin2004,Yuasa2004}. On top of this, it also has a structural inversion symmetry breaking at the interface (as the atomic lattice changes its periodicity and potential), which is the main underlying cause of the Rashba SOC~\cite{Zutic2004,Fabian2007}. In Figure~\ref{fig:MAAR-vs-H}a, we show how the conductance is only mildly suppressed if we compare it to what one could expect for a similarly spin-polarized interface ($\sim75\%$), which suggest that the spin-flip Andreev reflection process explained above is responsible for the relatively high values of $G(V=0)$ observed. 

Before establishing out conclusions, however, we have to take into account extrinsic phenomena which could produce similar results. The main contender in this regard is the possibility that superconducting vortices (vanadium is a type II superconductor) are being produced by the applied magnetic field, and since the V layer has a very pronounced aspect ratio (very thin compared to its lateral dimensions), this could affect differently to the conductance depending on the direction of the field, which could potentially ``mimic'' the conductance anisotropy. In particular, our V films are about 40~nm thick and a have a few tens of microns wide, which yields two critical fields: $H_{C2}=3.5$4~kOe for the out-of-plane direction, and $H_{C2}=12$~kOe in plane. To rule out this possibility, we performed control experiments that are shown in Figure~\ref{fig:MAAR-vs-H}b, where after an initial saturation of 0.3~T in the required direction (IP or OOP), we started to measure the zero bias conductance at $H=0$ and then slowly increased the field in the IP or OOP direction. These experiments clearly verify a zero bias conductance anisotropy induced by the non-volatile IP and OOP magnetization directions below the critical temperature (MAAR$\sim17\%$). Interestingly, looking at the $H=0$ data, the zero bias conductance shows that the superconducting gap is stronger in the out-of-plane configuration compared to the in-plane one. This is exactly the opposite to what we would observe if this difference was caused by superconducting vortices, since they would diminish the superconducting gap, \emph{increasing} the conductance~\cite{Martinez2020}.

To complete this picture, Figure~\ref{fig:MAAR-vs-H}e,f depict the zero bias conductance during an IP (0\degree) to OOP (90\degree) magnetization rotation, where the phenomenological model that was used to understand the results is also shown with solid lines. The reasonable fit allowed us to establish with greater certainty Rashba SOC as the main contributor to the MAAR\cite{Martinez2020}.

\subsection{Superconductivity-induced change in magnetic anisotropy}\label{section:MCA}

Due to the characteristic energies of the magnetic and superconducting order parameters, it is common to see from experimental results that the magnetism affects (or totally disrupts) superconductivity in a neighbouring material. The opposite is not common, since the magnetic exchange field in ferromagnets is more energetic than superconducting correlations ($\sim10^3$~K vs $\sim10^1$~K respectively). However, an important magnetic property of epitaxial materials, MCA does have a energy scale which is similar to the superconducting gap, and therefore it is more susceptible for a mutual interaction with superconductivity. This is precisely what we will show in this section, where the results let us observe how superconductity changes the easy magnetization axis of the adjacent ferromagnetic layer.

This phenomenon was known from theoretical studies since 2019~\cite{Johnsen2019}, but was not experimentally verified until a couple years later~\cite{gonzalez-prb-20,gonzalez-scirep-21}. The underlying physics involve Rashba SOC and LRT generation. In this case, in particular, the crucial part is that the generation rate of LRTs is sensitive to the relative direction of the magnetization and the SOC field. When the superconducting transition takes place, and to minimize the total energy in the whole system (V/MgO/Fe), the magnetization re-orients itself to a direction where \emph{less} LRTs are formed (as they are energetically expensive), even if that means aligning with the hard magnetization axis. Superconductivity acts as a ``switch'' for the magnetization, which could be a really useful approach for designing future devices where magnetization control is still a significant challenge to address~\cite{Banerjee2014,Baek2014,Gingrich2016,Satchell2021}.

For clarity, this section is separated in two parts, both studying the S/F/F junctions described in Figure~\ref{fig:samples}c. First, we will verify the superconductivity induced modification of the IP magnetic anisotropy, and after that we will discuss modification of the OOP anisotropy.

\subsubsection{Superconductivity-induced change in the IP magnetic anisotropy}\label{section:IP-MCA}

The most effective way of exploring the in-plane MCA landscape of the S/F/F junctions were magnetic field rotations, in which we used a field that was greater than the coercive field of the soft F layer but lower than the one of the hard F layer (see Figure~\ref{fig:magnetic-states}b). The range of 70 to 120~Oe worked best in our case, depending on the sample. Using this range of fields ensured that the hard F layer remained fixed, while the soft one could freely rotate its magnetization following the applied field direction, revealing its own MCA in the process. This is shown in Figure~\ref{fig:magnetic-states}b, where we can distinguish the two easy axes aligned with the [100] and [010] crystallographic directions as the soft F layer moves, setting the junction in the parallel P, perpendicular-in-plane (PIP) and AP magnetic configurations.

We used the experimental resistances to approximate the exact angle between both F layers in any situation where we could find different resistance values (found both above and below the critical temperature of vanadium). We used these resistances, for known magnetic configurations, to calibrate a modified version of the Slonczewski model~\cite{Slonczewski}, which can accurately model the S/F/F junctions due to their large effective spin polarization~\cite{Martinez2018,Parkin2004,Yuasa2004}.

\begin{figure}
\begin{center}
\includegraphics[width=\linewidth]{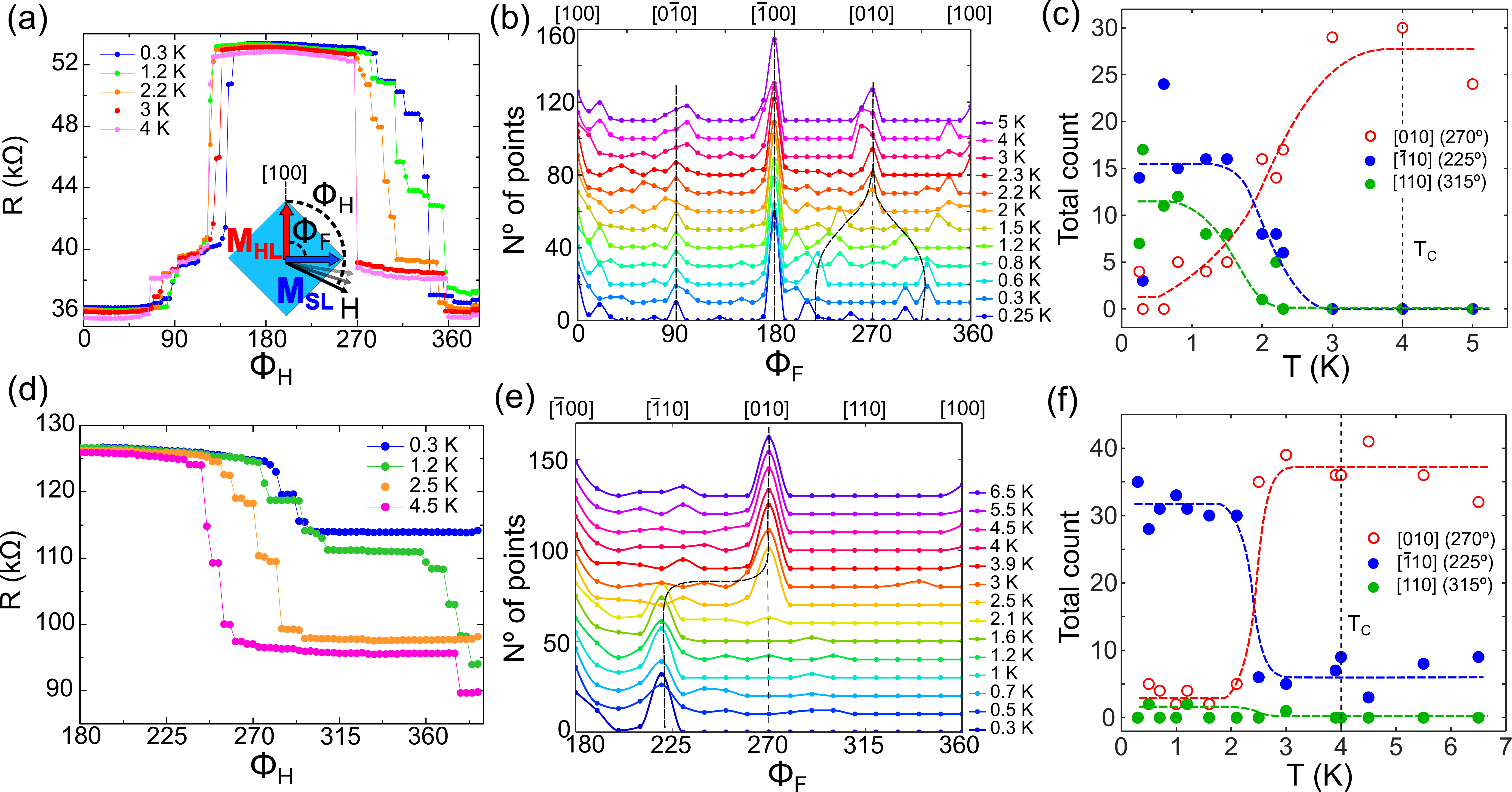}
\caption{Resistance vs in-plane magnetic field rotation angle for a V/MgO/Fe/MgO/Fe/Co junction, for different temperatures from above to below $T_C$, when the rotation is started from the P (a-c) and the AP (d-f) magnetic configurations, with the corresponding crystalline axis indicated. The inset helps visualizing the experiment, defining the angles between the two F layers ($\Phi_\text{F}$) and the one of the applied field ($\Phi_\text{H}$). (b) and (e) show 10$^\circ$ histograms from the data in (a) and (d) respectively. (c) and (f) take the data from (b) and (e) during the second half of the magnetic field rotation, representing each histogram's count vs temperature around the intermediate (normal state hard axis) states: $\text{AP}+45^\circ$ or the $[\overline{1}10]$ axis, $\text{AP}+90^\circ=\text{PIP}$ or [010], and $\text{AP}+135^\circ$ or [110]. Figure adapted from Ref. \cite{GonzalezRuano2020} with the necessary permissions.}
\label{fig:rotations}
\end{center}
\end{figure}

The magnetization rotations were then performed at different temperatures, from above to below $T_C$, in several samples. In Figure~\ref{fig:rotations}, we show above gap conductance versus applied magnetic field angle on two different samples, with the correspondent relative magnetization angle analysis. While the resulting TMR is not affected by temperature during the first half of the rotation, the second half has noticeable changes of the relative magnetization angle below $T_C$. In particular, we see new free layer magnetization configurations emerge in the superconducting state. According to the calibrated model, it seems to be oriented near the hard axes of the soft F layer, 45$^\circ$ from the easy directions. These new configurations disappear when increasing temperature and approaching $T_C$.
%%%%%%%%%%%%%

Before continuing to fit the experimental results to the theoretical model, we want to discard some extrinsic effects. In the IP situation, the main candidate is the formation of domain walls (DWs). To asses their possible role in changing the resistance of the junctions during the field rotations (which could potentially invalidate our macrospin approach with the Slonczewsky model), we used the MuMax3~\cite{mumax} micromagnetic simulation software to compare the formation of DWs in the [100] and [110] directions. The results showed that these DWs, in case they are present, should preferentially pin the magnetization along the [100] direction, and form less in the [110] one (see Ref.~\cite{gonzalez-prb-20}). Since we observe a high spin polarization when the magnetization is saturated along the [100] axis, indicating that no DWs are present, we can be reasonably confident that no DWs should appear during the rest of the rotations.

\begin{figure}
\begin{center}
\includegraphics[width=0.8\linewidth]{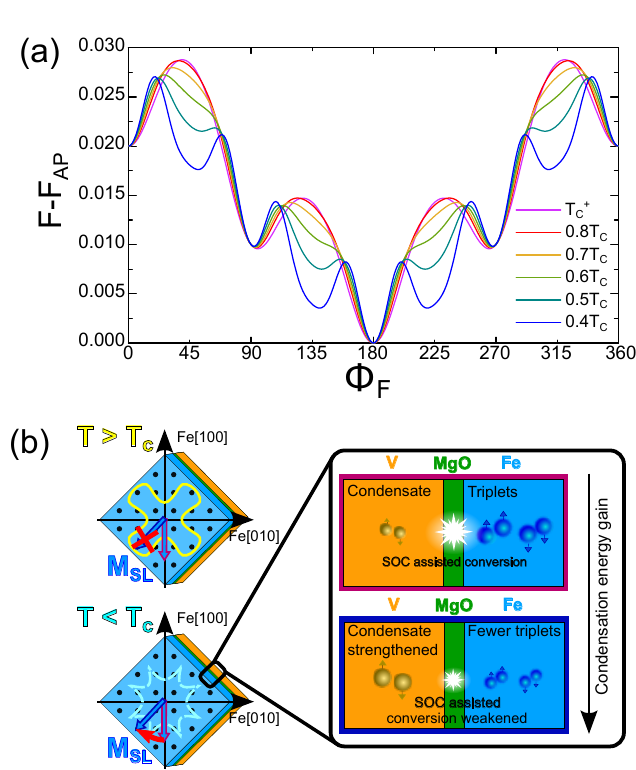}
\caption{(a) Free energy $F$ vs in-plane magnetization angle $\Phi_\text{F}$ for different temperatures around $T_C$. (b) Sketch of the physical mechanism behind the S-induced change in magnetocrystalline anisotropy (MCA): Above $T_C$, V is a normal metal, while the Fe layer has a 4-fold in-plane MCA (yellow line). Below $T_C$, V is superconducting and influences the Fe with a magnetization-dependent penetration of Cooper pairs. The generation of triplets is lower for the $[\overline{1}\overline{1}0]$ direction, maximizing the superconducting condensate energy gain. This in turn modifies the effective MCA of the Fe layer (cyan line), allowing the magnetization to reorient towards the $[\overline{1}\overline{1}0]$ axis (blue arrow). Figure adapted from ref. \cite{GonzalezRuano2020} with the necessary permissions.}
\label{fig:IP-modelling}
\end{center}
\end{figure}

Once the extrinsic factors were taken care of, we turn to the theoretical model~\cite{Johnsen2019}, adapting it to our system. As mentioned, the main idea is that the invariance of the superconducting proximity effect to the angle of the F layer magnetization is broken in the presence of SOC~\cite{Jacobsen2015,Johnsen2019}. Formally, the model is based on a tight-binding Bogolioubov-de Gennes Hamiltonian, which includes terms for electron hopping both within and between the different layers, interfacial Rashba SOC, spin exchange splitting for the Fe layers, and conventional superconductivity in vanadium. The total free energy is calculated by taking into account the magnetization-dependent superconducting proximity effect, bi-axial in-plane MCA, and a weak antiferromagnetic coupling between the F layers that was observed experimentally. The result is shown in Figure~\ref{fig:IP-modelling}, where the free energy is plotted vs the angle of the magnetization ($\Phi_\text{F}$) under different temperatures. This nicely reproduces the new 45$^\circ$ easy axis below $T_C$. For more calculation details, see\cite{GonzalezRuano2020}.

\subsubsection{Superconductivity-assisted change of the OOP magnetic anisotropy}\label{section:OOP-MCA}

We turn now to describe superconductivity-induced changes in perpendicular magnetic anisotropy of V/MgO/Fe/MgO/Fe/Co junctions. Precise control over the OOP or PMA in thin ferromagnetic films allowed to further reduce the size of magnetic bits when they were already reaching a critical size~\cite{perpendicular,Dieny2017}, although fine tuning it also became a difficult challenge~\cite{Chuang2019,Yi2021,Sun2017}. The main strategy to achieve an OOP magnetic orientation has usually been varying the F layer thickness and adding oxide barriers, but there is also another alternative: temperature or electromagnetic pulses (typicallly in the microwave frequency range) can decrease the energy barrier between the IP and OOP configurations, and then an appropiate magnetic field can trigger the desired reorientation~\cite{Challener2009,Zhu2008,Martinez2018}.

\begin{figure}
\begin{center}
\includegraphics[width=\linewidth]{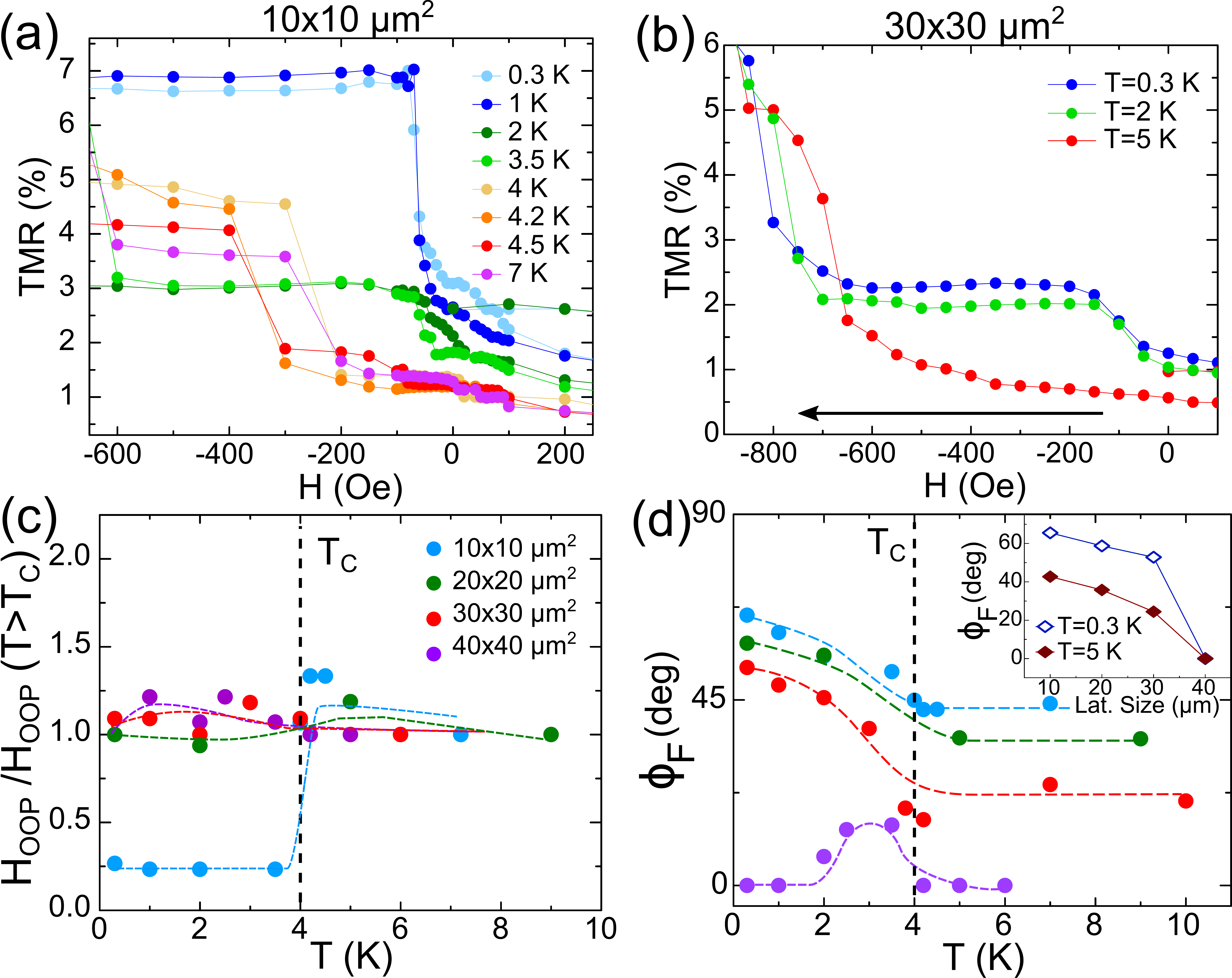}
\caption{(a) Out-of-plane (OOP) reorientation in a $10\times10$~$\mu\text{m}^2$ S/F/F junction, for different temperatures around $T_C$. (b) The same type of experiment for a larger ($30\times30$~$\mu\text{m}^2$) junction, where an incomplete reorientation can be inferred from the TMR. (c) $H_{\text{OOP}}$ vs $T$ in different lateral sized junctions. A strong reduction of $H_{\text{OOP}}$ takes place below $T_c$ for the junction of size $10\times10$~$\mu\text{m}^2$. (d) $\Phi_\text{F}$ vs $T$ at $H=0$ for four different sized junction (same legend that in the previous panel). The inset compares the angle at $H=0$ and $T=5$~K (above $T_C$) and at $T=0.3$~K (below $T_C$) for the different lateral sizes. The colored dashed lines are guides for the eyes, while the vertical, black, dashed lines mark $T_C$. Figure adapted from ref. \cite{GonzalezRuano2021} with the necessary permissions.}
\label{fig:zero-field-OOP}
\end{center}
\end{figure}

In the case of our S/F junctions, all the necessary ingredients are in place to experimentally achieve the prediction of the modification of the MCA below $T_C$ for the OOP case~\cite{Johnsen2019}. Furthermore, we have studied several different samples with varying lateral sizes, which (on top of reproducibility) allows for different relative weights of the shape (IP) and surface (OOP) magnetic anisotropies. In the case of the smallest junctions shown in Figure~\ref{fig:zero-field-OOP}, there is a strong competition between the two anisotropies, which enabled the observation of the full OOP reorientation predicted in Ref.\cite{Johnsen2019}. Even for larger junctions where the IP component dominates, we still observed weaker but robust changes suggesting a higher relative strength of the OOP anisotropy.

In Figure~\ref{fig:zero-field-OOP}, it can be seen how the OOP reorientation changes under different temperatures for a $10\times10$~$\mu\text{m}^2$ junction. Below $T_C$, there is a decreasing in the applied field that triggers a full OOP magnetization reorientation ($H_{\text{OOP}}$). In junctions with larger lateral dimensions, the same type of reorientation was observed below $T_C$, althoughthe effect was weaker (Figure~\ref{fig:zero-field-OOP}b,c). This behavior is consistent with an incomplete OOP reorientation or the reorientation of only certain magnetic domains. As in the IP case, the angle between the two F layers is shown in Figure~\ref{fig:zero-field-OOP}d.

It is worth noting that the reason for the OOP TMR experiments being ``asymmetric'' was an already known phenomenon~\cite{Martinez2018}, explained by a different density of lattice dislocations between the top and bottom parts of the Fe layer nearest to the V one. This behavior is inherent to most epitaxial fabrication methods, and in our case results in slightly different surface anisotropies, which in turn means that the magnetization is reoriented OOP preferentially for one of the field directions compared to the other one.

As in the IP case, we also carried out micromagnetic simulations using MuMax3, trying to understand the possible influence of defects in the magnetization (preferentially near the surfaces of the Fe layer). We modelled~\cite{GonzalezRuano2021} these defects as cells with a higher saturation magnetization ($M_S(\text{defects})=1.25\times M_S(\text{Fe})$), finding out that a density of $\sim10^{-3}\%$) does not change the non-volatility of the OOP magnetic configuration, although it does affect the reorientation field to the IP configuration $H_\text{OOP-IP}$. If the defect density is further increased up to $\sim2\times 10^{-3}\%$, this transition becomes volatile, which is not what we observe experimentally. This is an indication of the low density of magnetic defects in the junctions.

The theoretical model \cite{GonzalezRuano2021} is also very similar to the one used to model the changes in the IP MCA. The free energy is computed from a Bogoliubov-de Gennes (BdG) Hamiltonian, showing that triplet generation has an energy cost that triggers the OOP reorientation below $T_C$. The Hamiltonian also includes a preferred IP MCA with cubic symmetry and the superconducting proximity effect. Similarly to the IP case, a new local energy minimum appears for the OOP magnetization when the temperature is lowered enough.

\begin{figure}
\begin{center}
\includegraphics[width=\linewidth]{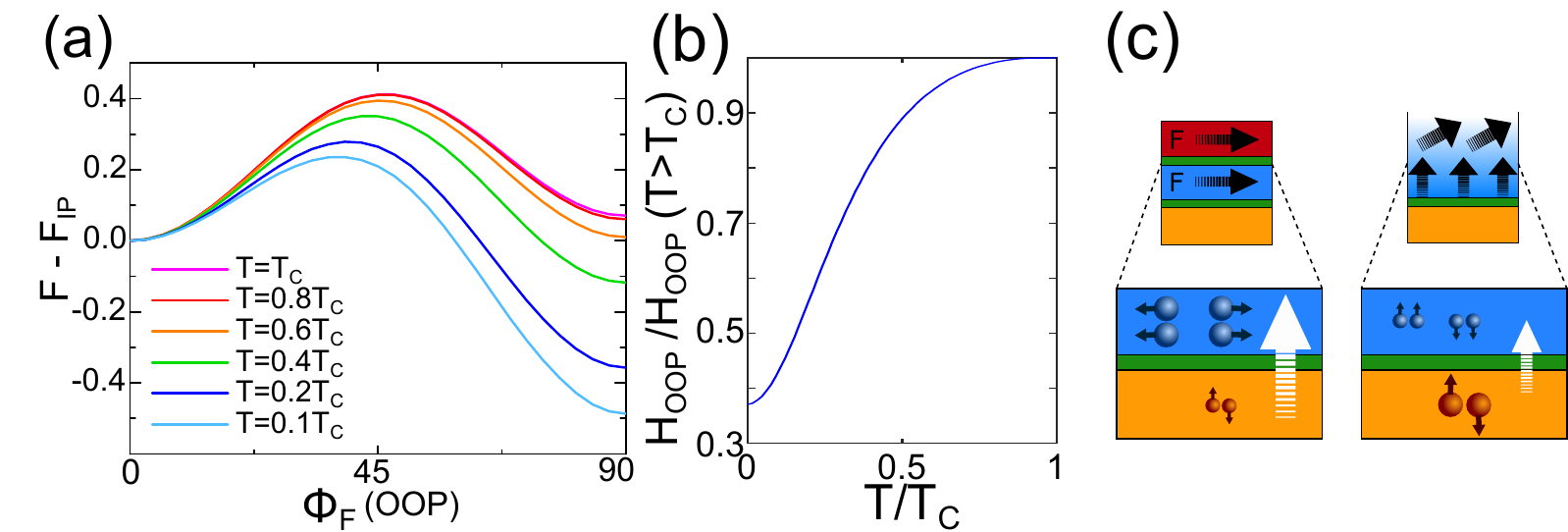}
\caption{(a) Free energy ($F$) vs out-of-plane (OOP) angle $\Phi_\text{F}$, compared to the free energy in the in-plane (IP) configuration ($F_\text{IP}$), for temperatures around $T_C$. (b) The OOP minimum becomes deeper, explaining the observed decrease of $H_{\text{OOP}}$. (c) When the soft F layer reorients to an OOP alignment, the spin-orbit coupling-mediated transformation of singlet Cooper pairs into long-range triplets is less efficient, which makes the superconducting condensate stronger, in turn decreasing the OOP free energy as $T$ goes below $T_C$. Figure adapted from ref. \cite{GonzalezRuano2021} with the necessary permissions.}
\label{fig:BdG-theory-OOP}
\end{center}
\end{figure}

The result of the modelling is shown in Figure~\ref{fig:BdG-theory-OOP}, where the free energy is plotted vs IP-OOP angle for different temperatures. Below $T_C$, the free energy minimum for the OOP configuration (which is already present due to the introduction of the PMA in the model, but it is not the ground state above $T_C$) gets lower, explaining how it becomes a much easier magnetization axis that in turn helps lower the field necessary to induce the OOP transition, and makes it robust under zero field.

\subsection{Conductance anomalies: disentangling SOC and spin texture contributions to LRTs formation}\label{section:CAs}

For a long time, the conversion of ``conventional'' singlet Cooper pairs to equal-spin triplet pairs in S/F interfaces has been linked to inhomogeneous magnetism or spin textures (STs)~\cite{uti1999,Bergeret2001,keizer-nature-06,Khaire2010,Robinson2010}. Theoretical works~\cite{Gorkov2001,Bergeret2013,Bergeret2014,Jacobsen2015} showed that LRT pair formation could also be generated by interfacial SOC, which as we have seen was later confirmed experimentally~\cite{Banerjee2018,Martinez2020,GonzalezRuano2020,GonzalezRuano2021,Jeon2020,Cai2021}.

\begin{figure}
\begin{center}
\includegraphics[width=\linewidth]{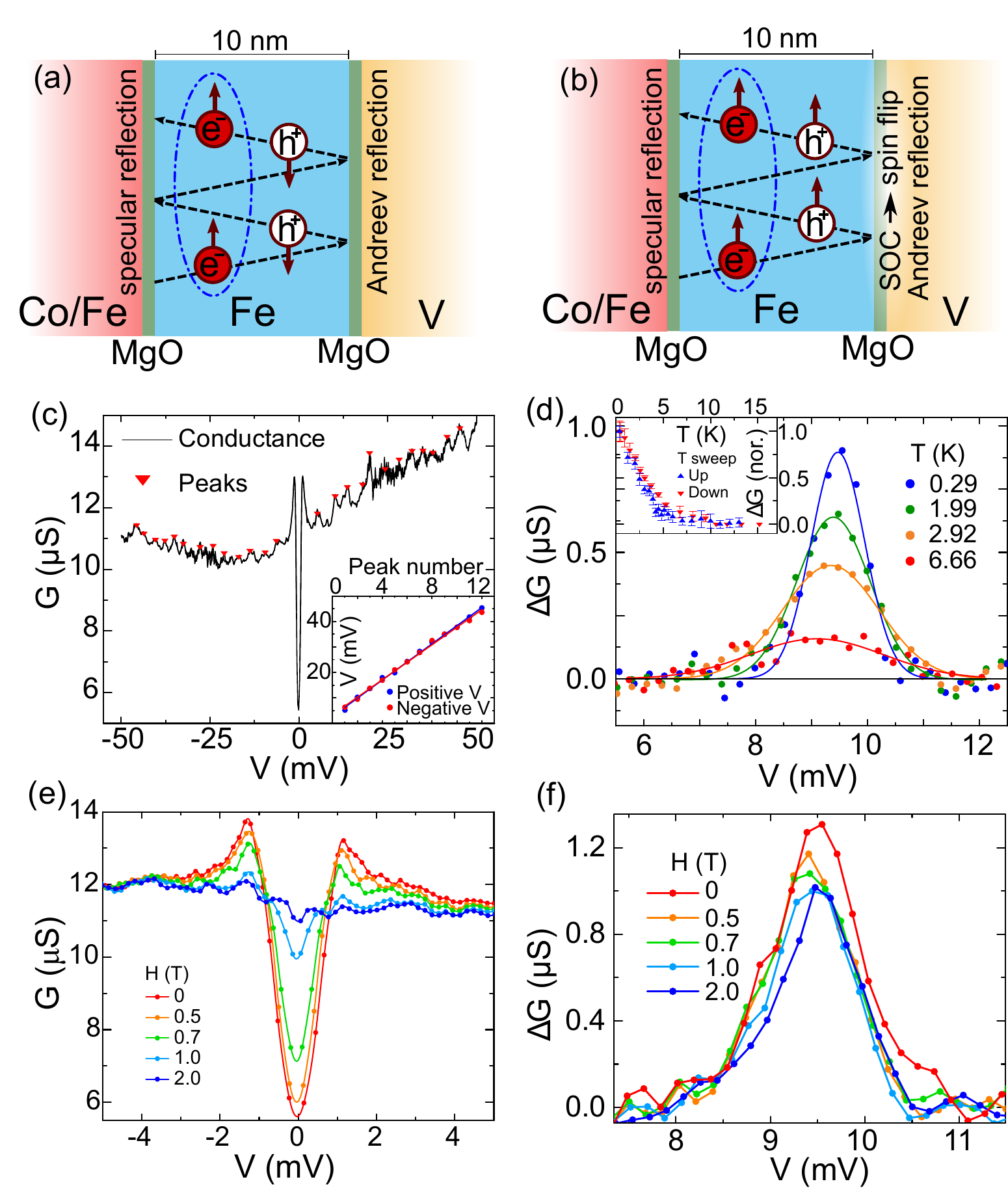}
\caption{(a)~Conventional and (b)~spin-flip McMillan-Rowell resonances due to interfacial spin-orbit coupling. An electron in the F layer (lower part) undergoes Andreev reflection at the interface. The interference takes place between the electrons circled in blue, producing the conductance anomalies (CAs). (c)~CAs in a $20\times20~\mu\text{m}^2$ S/F/F sample. Inset: periodicity analysis, showing peak bias vs peak number for the positive (blue) and negative (red) bias ranges. (d)~Amplitude above baseline conductance of a CA ($\Delta$G) vs $T$, from below to above $T_C$ (blue in the inset) and vice-versa (red). (e)~Conductance vs applied bias around the superconducting gap for different in-plane (IP) magnetic fields. (f)~Same peak analyzed in (d) under high IP applied fields above the one suppressing the superconducting gap. Adapted from Ref.\cite{Tuero2024} with the necessary permissions.}
\label{SFF-CAs}
\end{center}
\end{figure}

It is well known that in clean N/F and S/F hybrids, Fabry-P\'erot interferences due to electron confinement can result in the so-called MacMillan-Rowell resonances (MRRs)~\cite{MRR1,MRR2}, originated as two Andreev and two normal reflections taking place inside the F electrode interfacing the S layer~\cite{deJong1995,Visani2012,Melnikov2012,Costa2022} (Figure~\ref{SFF-CAs}a). Tomasch resonances~\cite{Tomasch1966} (TRs) are similar phenomena occurring in the S layer that have been recently put forward to distinguish between conventional and topological superconductivity~\cite{Strkalj2024}. The above-gap conductance anomalies (CAs) arising from these two processes were suggested as experimental proof for LRTs formation in S/F heterostructures~\cite{Visani2012}. The full suppression of these CAs under IP magnetic fields pointed towards a direct link between LRT generation and spin textures~\cite{Visani2015}.

In this section, we demonstrate that studying MRR-induced CAs in S/F junctions under IP and OOP applied fields provides a unique tool to disentangle the STs and SOC contributions to LRTs. The analysis is based on the consideration that SOC-induced LRTs (SOC-LRTs) should be more robust to magnetic field than those induced by spin textures (ST-LRTs), which should strongly diminish under a high field saturating the magnetization. As we demonstrate below, the CAs remain mostly unaffected under high applied fields, firmly pointing towards SOC-LRTs. However, at low magnetic fields (where more STs appear) the CAs amplitude strongly depends on the field direction. The results are qualitatively supported by micromagnetic simulations pointing towards the anisotropic induction of spin textures in the Fe layer under low IP and OOP magnetic fields. 

Figure~\ref{SFF-CAs}a,b illustrates the MacMillan reflections in the absence (a) or presence (b) of spin flips, induced by interfacial STs and/or SOC. Figure~\ref{SFF-CAs}c shows that the conductance of S/F/F junctions measured in P and AP states (at $T=0.3$~K$=0.075~T_C$ unless otherwise stated) present several above-gap CAs which have a periodicity with applied bias (see inset in Figure~\ref{SFF-CAs}c). As mentioned above, this kind of CAs have been linked to MacMillan and Tomasch reflections~\cite{Visani2012,Martinez2020} generating LRTs in S/F systems with highly polarized ferromagnets. The periodicity suggests that MRRs are the leading interference mechanism, since these reflections give rise to conductance oscillations at periodic characteristic bias:

\begin{center}
  \begin{equation}\label{eq:peaks}
    V_n=V_0+nhv_F^\text{Fe}/4t_\text{Fe},
  \end{equation}
\end{center}

where $n=0,1,2,...$ label the successive conductance oscillation peaks, $h$ is the Planck constant, $v_F^\text{Fe}$ is the Fermi velocity in the layer where the interference takes place (in our case, the soft Fe layer), and $t_\text{Fe}$ its thickness (10~nm in our S/F/F junctions and 45~nm in the F/S/F ones). For MRRs to occur, the gap-induced phase coherence between the initial electron and the reflected hole is maintained through the whole path back and forth in the F layer (a total distance of at least 40~nm and 180~nm for the S/F/F and F/S/F junctions, respectively). This is a strong hint of long-range superconducting proximity effects, as it means that superconducting correlations must survive deep into the Fe layer.

Figure~\ref{SFF-CAs}d shows how the CAs emerge mainly below $T_C$. 
Both the sub-gap conductance and one of the CAs in an S/F/F junction are studied as a function of magnetic field in Figure~\ref{SFF-CAs}e-f. The superconducting gap is suppressed by an IP magnetic field of about 2~T, while the CAs remain almost unaffected by the same field. Nevertheless, we note that the second critical field was estimated from experiments that probe the transport through the V/MgO/Fe interface, where the tunneling current is sensitive to the superconducting gap formed within the first few V/MgO interfacial atomic layers (where spin triplets could also be generated via SOC). In principle, the interfacial layers may host gapless states, whereas the inner layers— deeper within the 40-nm-thick vanadium and not accessed in our experiments— could remain fully gapped.

The experiments leads us to two conclusions. First, the CAs are linked with superconductivity, since they disappear above $T_C$. Second, they are related to LRTs, which can survive high applied fields and inside the ferromagnet's exchange field. We exclude resonant transmission of normal electrons through quantum well states~\cite{Tao2015, Tao2019} as the main mechanism behind the observed CAs (see Ref.\cite{Tuero2024} discussion and its Supplemental Material for more details on this). Overall, the observation of CAs is indicative of the smoothness of the interfaces in the junctions ensured by the layer-by-layer epitaxial growth, which allows for the observation of quasiparticle interference effects. 

\begin{figure}
\begin{center}
\includegraphics[width=\linewidth]{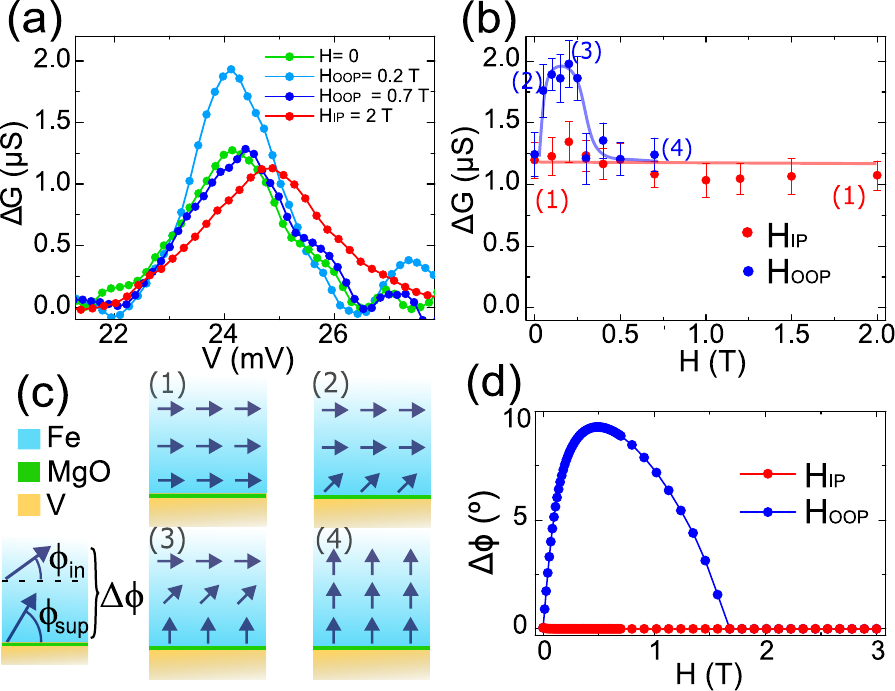}
\caption{(a)~$\Delta G(V)$ of a conductance anomaly (CA) for different in-plane (IP) and out-of-plane (OOP) applied fields in a $30\times30~\mu\text{m}^2$ S/F/F junction over the background conductance. (b)~Amplitude of the CA vs IP and OOP fields. Error bars mark the 95\% confidence bounds. The lines are guides for the eye. The numbers within the graph correspond to the steps in panel (c), where we sketch the magnetization configuration and define the magnetization gradient (spin texture) $\Delta\phi$ simulated in the next panel. (1)~After an initial IP saturation of 0.3~T, the magnetization is homogeneous. (2)~A small OOP field starts inducing an angle gradient near the surface due to the perpendicular magnetic anisotropy. (3)~as the OOP field increases, the angle gradient reaches a maximum. (4)~a large enough OOP field saturates the magnetization, returning to a reduced spin gradient. (d)~Dependence of the spin gradient $\Delta\phi$ vs IP and OOP applied field from micromagnetic simulations. Adapted from Ref.\cite{Tuero2024} with the necessary permissions.}
\label{CAs-vs-H}
\end{center}
\end{figure}

To investigate the possible contribution of spin textures to the CAs and LRT formation, we have studied them in a field range not yet saturating the Fe layer magnetization. Figure~\ref{CAs-vs-H}a shows a $G(V)$ curve around a CA peak of a S/F/F junction at $T=0.3$~K after the background conductance is subtracted. This CA persisted for fields up to $H_\text{IP}=2$~T and $H_\text{OOP}=0.7$~T. The dependence of the CA with IP and OOP applied fields is presented in Figure~\ref{CAs-vs-H}b. It remains approximately constant for an IP magnetic field up to 2~T, in qualitative agreement with experiments on smaller junctions. When an OOP field is applied, the CA is first enhanced with respect to the base level. The maximum is reached for $H_\text{OOP}\approx0.3$~T, and further increasing the OOP field removes the CA excess (see Figure~\ref{CAs-vs-H}b). 

To understand these results, we consider the role of PMA-induced spin textures at the MgO/Fe interface, sketched in Figure~\ref{CAs-vs-H}c. The CA amplitude under IP field gives a baseline for the spin-triplet generation rate via SOC, as the magnetization of the Fe layer in this situation is saturated thanks to the IP anisotropy. When a low $H_\text{OOP}$ is applied, the Fe atomic layers near the surface align with this field due to the PMA, increasing the angle $\phi$ with the IP direction. The presence of these spin textures close to the MgO/Fe interface could then open an additional channel for LRT generation~\cite{Bergeret2005,Robinson2010,Buzdin2007}, which would enhance the CA. When the OOP field is further increased, the inner layers of Fe (which were initially oriented IP) would also align with the OOP magnetic field, removing the STs (therefore reducing the angle gradient $\Delta\phi$) and letting the V/MgO/Fe system only with the SOC contribution to LRT formation.

This scenario is supported by micromagnetic simulations (Figure~\ref{CAs-vs-H}d) showing that, while an IP field or large OOP field quickly saturates the magnetization, a moderate OOP field can initially maximize the angle between the magnetization of the atomic layers interfacing the MgO where PMA is present, and the inner layers where the IP shape anisotropy dominates ($\Delta\phi$ in Figure~\ref{CAs-vs-H}d, the definition is sketched in panel c).

\begin{figure}
\begin{center}
\includegraphics[width=\linewidth]{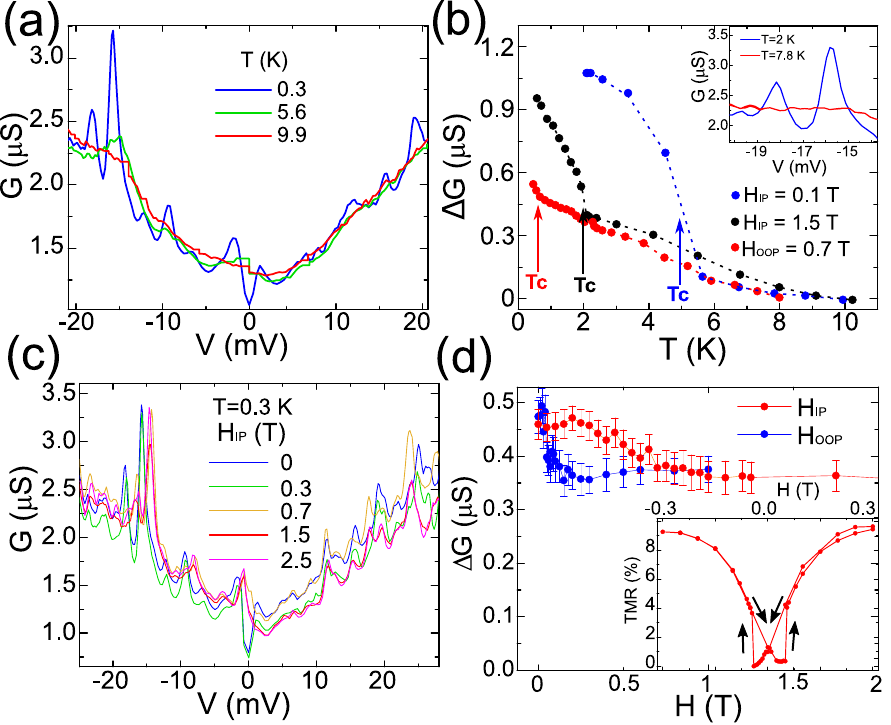}
\caption{Conductance anomalies (CAs) in a F/S/F junction. (a)~Conductance vs bias below (blue), near (green), and above (red) $T_C$, showing how the CAs disappear in the normal state. (b)~CA amplitude near $-15$~mV vs $T$ for different applied fields, in-plane (IP) and out-of-plane (OOP). The lines are guides for the eyes, and the arrows indicate the estimated $T_C$ of each experiment. The inset shows the conductance around the CA below and above $T_C$. (c)~Conductance vs bias of the same junction at $T=0.3$~K for different IP applied fields, showing the robustness of the CAs under large magnetic fields. (d)~amplitude of a CA vs applied field, both IP and OOP. The inset shows an IP magnetoresistance measured at $T=0.3$~K, showing that the IP magnetization does not fully saturate until an applied field of about 0.3~T. Adapted from Ref.\cite{Tuero2024} with the necessary permissions.}
\label{sandwich}
\end{center}
\end{figure}

Similar CAs were also observed in F/S/F junctions (the structure of these samples is sketched in Figure~\ref{fig:samples}d, and the data is shown in Figure~\ref{sandwich}a). They are also robust under high applied magnetic fields, and show a sharp transition near $T_C$ (Figure~\ref{sandwich}b,c). However, as shown in Figure~\ref{sandwich}d, the behavior of CAs with IP and OOP fields is in some way different to one observed in S/F/F junctions. While in both types of junctions the CAs maintain a baseline level in the high field regime, in the F/S/F samples the CA enhancement at low fields is seen both for IP and OOP fields. This is the result of the F/S/F junctions having a much stronger AP coupling compared to the S/F/F ones, which could induce additional STs in the low IP field regime. Combined with the PMA, this results in the magnetization only becoming fully saturated under an IP applied field above $0.3$~T (see Figure~\ref{sandwich}d inset, showing an IP magnetoresistance). In contrast, this extra amplitude of the CAs disappears more quickly with an OOP applied field, as in this case the PMA and the applied field point in the same direction. The CAs still persist in the high field regime (Figure~\ref{sandwich}c,d) supporting a scenario of dominant SOC-induced LRT pair generation with an additional contribution of ST-LRTs in the low field regime.

\section{Excess subgap shot noise in S/F junctions}\label{section:SN}

Electronic voltage fluctuations (also known as noise), and in particular, \emph{shot noise}, can be a powerful tool to extract otherwise unaccessible information about the process of electronic transport, which can not be inferred from the conductance~\cite{Etienne1997,Blanter2000,Aliev2018Book}. This kind of measurements have already shown their potential in superconductivity, being used to reveal details such as electron pairing in pseudogap states~\cite{Panpan2019}, the symmetries of quasi-particle pairing~\cite{Benjamin2019} or their charge~\cite{Ronen2016}. In our S/F junctions, we have used the shot noise to inspect the transport both in the normal and superconducting regimes, for biases above and below the superconducting gap. When superconductivity appears, we consistently observed shot noise values that exceeded the expected values for Cooper pair transport in more than an order of magnitude~\cite{GonzalezRuano2025}. In order to understand this unforeseen effect, a simplified model has been developed that links the observations with electron bunching caused by multiple spin-flip Andreev reflections, happening thanks to the presence of interfacial SOC~\cite{GonzalezRuano2025}.

\begin{figure}
\begin{center}
\includegraphics[width=\linewidth]{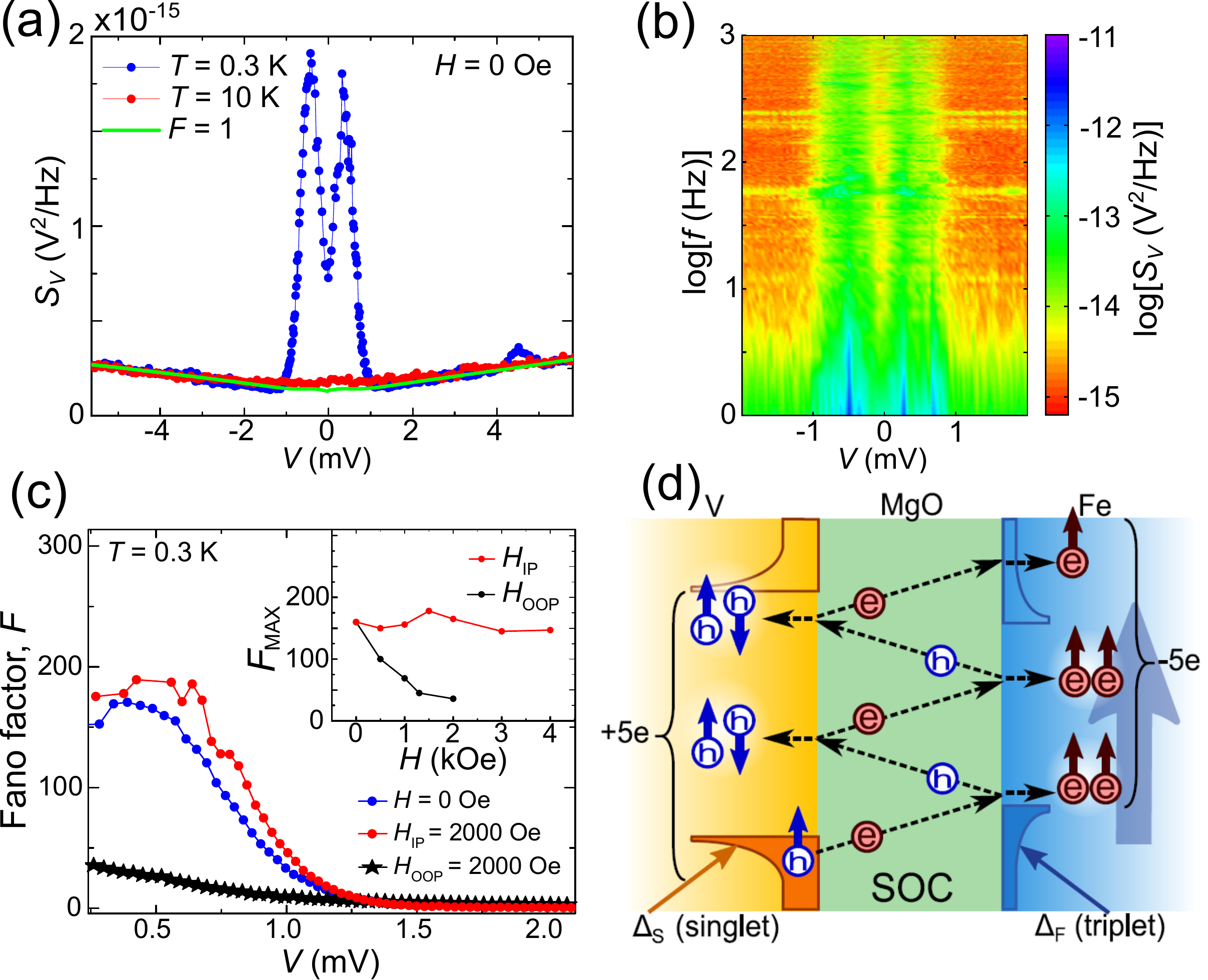}
\caption{(a) Shot noise for a S/F junction at $T=10$~K and $T=0.3$~K. Above the gap, the noise level increases with bias. Below the gap, the noise level increases more than an order of magnitude, with two distinct peaks. The solid green line corresponds to a Fano factor of 1. (b) noise vs frequency vs bias, showing that the frequency response of the noise spectrum is flat in a wide frequency range, with $1/f$ contribution only below 10~Hz. (c) Fano factor vs different applied fields. (d) tentative model of the multiple Andreev reflection process that could cause the observed behavior. Adapted from Ref.\cite{GonzalezRuano2025} with the necessary permissions.}
\label{fig:SN}
\end{center}
\end{figure}

In all of the S/F junctions that we studied, for temperatures or biases above the superconducting gap, the dominant component of the noise is Poissonian, as shown in Figure~\ref{fig:SN}a. The noise excess only appears below $T_C$ and in the superconducting gap region. The voltage noise power is frequency independent for a frequency range of more than two orders of magnitude (Figure~\ref{fig:SN}b). The excess noise gives us a subsequent increase of the associated Fano factor. Shown in panel Figure~\ref{fig:SN}c, the Fano factor compares the noise measured experimentally with what one would expect from a purely Poissonian noise source: a voltage shot noise of $S_V(V)=F\cdot S_I\cdot(dV/dI)^2=F\cdot 2eIR^2$, where $S_I=2eI$ is the Poisoninan current noise for Cooper pairs (charge$=2e$). Following the noise signal, the Fano factor increases from a normal state value close to 1 to $\sim10^2$ inside the superconducting gap. It reaches its maximum values at the lowest achievable temperatures ($\sim0.3$~K in our cryostat), decreasing when the junction is heated up towards $T_C$ or under applied magnetic fields (Figure~\ref{fig:SN}c).

We have verified if superconducting vortices, which could be generated by the stray field of the F layer near the superconductor or applied fields, could be responsible for the observed enhanced sub-gap shot noise. These vortices have a distinct noise signal in the frequency range, showing a $1/f$ dependence of the noise power with frequency~\cite{Anton2013}. However, we analyzed the noise spectrum and found that it is nearly frequency independent. This analysis is shown in Figure~\ref{fig:SN}b, revealing a $1/f$ component in the signal below 10~Hz. Furthermore, it doesnot increase with an OOP magnetization alignment or external magnetic field, which would produce more superconducting vortices. Detailed analysis dismissing the influence of superconducting vortices may be found in Ref.\cite{GonzalezRuano2025}.

Another posible explanation could be based in previous results~\cite{Wei2010}, which showed that strong low-bias conductance anomalies could result in corresponding anomalies in shot noise signal. If this were the case, our model predicts that the shot noise signal in our junctions would increase by at most a factor of 1.4, whereas the experiments show a $10-100$ times increase.
  
The next step is to turn our attention to an intrinsisc phenomenon: giant shot noise was found in S/S~\cite{Dieleman1997} and S/N/S~\cite{Hoss2000} junctions and explained with multiple Andreev reflections (MAR)~\cite{Averin1996,Cuevas1999}. The advantage of using MAR to expalain the observed shot noise resides in that (due to bunching) MAR transfers a larger effective charge, which can result in huge subgap shot noise at low biases. However, to the best of our knowledge, such effect has not been reported so far in devices with only a single superconducting layer. Taking this into account, the model used is based on the possibility of the inducing (through proximity effect) of a second interfacial superconducting region in the junction inside the F layer~\cite{GonzalezRuano2025}. Then, SOC could allow spin-flip Andreev reflections to form LRTs. This is basically what happens in Josephson junctions (JJs)~\cite{Octavio1983,Dieleman1997}, where noise excess well above the Poissonian levels have been reported and explained by multiple Andreev reflections. A proximity-induced gap in a ferromagnet near a superconductor was also predicted in previous works~\cite{Huertas-Hernando2005,Golubov2012}. Similar phenomena was also suggested to explain experimental data resembling Josephson effects in S/N structures~\cite{Han1990}. Figure~\ref{fig:SN}d sketches our proposed model. In the model, $e-h$ quasi-particles with $\Delta_1$ symmetry are available near the Fermi level of the Fe electrode, and could be transferred through the MgO~\cite{Butler2001}. The $e-h$ pairs would then be reflected multiple times before finally penetrating in the opposite electrode, producing the huge increase of the shot noise signal in the superconducting regime.

While the above simple model~\cite{GonzalezRuano2025} describes the increase of the below gap shot noise, it provides only a moderate increase (up to 10 times) of the Fano factor below $T_c$, while the experimentally observed giant shot noise values result in a Fano factor exceeding $F=100-200$. Therefore, the focus on the simple and transparent theoretical approach \cite{GonzalezRuano2025} invites future theoretical extensions involving also the influence of the above discussed SOC bottleneck mechanism emerging from the orbital symmetry mismatch between the superconducting and ferromagnetic electrodes. Indeed, both Rashba and Dresselhaus SOC involve random scattering processes \cite{Egues2002} (the main origin of decoherence of solid state qubits \cite{Bermeister2014}) which could provide an additional contribution to the giant shot noise missed by the minimal model \cite{GonzalezRuano2025}.

\section{Josephson Junctions based on S/SOC/F systems}\label{section:JJs}

One of the key potential applications of spin-triplet superconductivity lies in JJs, which serve as fundamental building blocks for qubits and superconducting spintronics in general~\cite{Dartiailh2021,Cai2022}. Traditionally, the control of spin supercurrrents in such devices has relied on magnetic textures~\cite{Junxiang2024}. However, once the role of interfacial SOC in generating LRTs has been the demonstrated experimentally \cite{Martinez2020,GonzalezRuano2020,GonzalezRuano2021,Tuero2024} a new pathway for the SOC control over spin-supercurrents has emerged. Although for vertical (pillar-type) S/F geometries the maximum values of the long-range spin triplet component are predicted in the presence of a non-volatile, OOP magnetization component \cite{Bergeret2014}, in practice, such an out-of-plane magnetization could induce superconducting vortices, complicating the interpretation of the results. Indeed, earlier investigations of the vertical Josephson junctions incorporating SOC have yielded inconclusive results\cite{Satchell2019}. Lateral geometries are advantageous because they impose fewer constraints on the generation of LRTs compared to stacked geometries, where the magnetization must lie between IP and OOP to optimize the spin-supercurrent generation\cite{Jacobsen2015}.

To enable a deeper investigation of the Josephson effect in superconductor/ferromagnet/superconductor (S/F/S) junctions with SOC, two theoretical studies proposed exploring the generation of LRTs in lateral JJs through IP magnetization rotations ~\cite{Bujnowski2019,eskiltprb2019}. The core concept behind the magnetic control of spin supercurrents is that both the sign and magnitude of the critical current are governed by spin precession and anisotropic spin relaxation\cite{Bujnowski2019}, which are highly sensitive to the direction of the applied field. As a result, the LRTs in relatively long ferromagnetic bridges decay weakly along the magnetization direction. However, when the exchange field is orthogonal to the spin diffusion direction, the spin supercurrent not only becomes significantly smaller but also exhibits oscillations between the SOC/S contacts\cite{eskiltprb2019}. Interestingly, the first experimental evidence of long-range Josephson effect was reported in lateral junctions based on the half-metallic ferromagnet CrO$_2$\cite{keizer-nature-06}, although the nature of the spin-active interface remained unclear. Later, the generation of LRTs in such structures was attributed to SOC at the S/F contact regions\cite{Bergeret2014}, although no studies at the time examined its dependence on the magnetization direction within the F layer.

%\setkeys{Gin}{draft=false}
\begin{figure}
\begin{center}
\includegraphics[width=\linewidth]{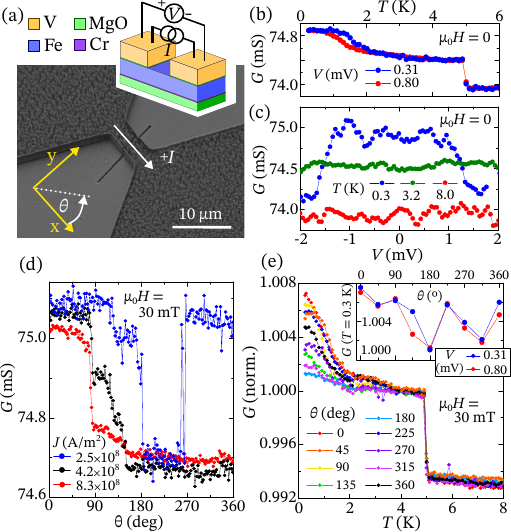}
\caption{(a)~SEM image of the lateral junctions under study with the zone attacked by focus ion beam (FIB) in the center. The yellow arrows and the angle $\theta$ define the coordinate system used for the external magnetic field. The white arrow defines the sign of the current with respect to the coordinate system. In the inset, a sketch of vertical structure of the junctions under study and the electronic set-up for the measurements. (b)~Conductance versus temperature $T$ with $\mu_0H=0$ measured at two different biases. (c)~Differential conductance versus bias curves at $\mu_0H=0$ for different temperatures. (d)~Conductance versus angle of the applied in-plane external magnetic field $\theta$ for different current densities across the junction, that correspond with biases of about 1.20~mV, 0.54~mV and 0.31~mV, from highest to lowest. (e)~Conductance versus temperature curves for different $\theta$. In the inset, conductance at $T=0.3$~K as a function of $\theta$.}
\label{fig-JJs}
\end{center}
\end{figure}
%\setkeys{Gin}{draft}

In the following part of this section we cover some of the preliminary results obtained in our investigation of magnetic field dependent electron transport in lateral JJs based on V/MgO/Fe epitaxial system. 
In order to achieve the S/SOC/F/SOC/S junctions in a lateral geometry along Fe (100) direction, a Fe(40~nm)/MgO(2.1~nm)/V(40~nm) epitaxial sample was grown with methods similar to those explained in Section~\ref{section:Growth}. Then, the sample was litographed into a $\sim3~\mu\text{m}$ wide bridge between two larger contacts shown in Figure~\ref{fig-JJs}a. Finally, the vanadium part of the bridge was cut across using focus ion beam (FIB) to remove (partially or fully) a strip of vanadium in a $\sim$250~nm wide region, achieving two separate V islands connected by a Fe bridge below them. At this stage, we monitored the FIB process by measuring the bridge resistance after consecutive and gradual FIB cuts to avoid cuting into the Fe layer, preserving the V-MgO-Fe-MgO-V current path. Both the SEM image and a sketch of the resulting lateral junctions and the electrical set up to perform transport measurements are shown in Figure~\ref{fig-JJs}a.

Figure~\ref{fig-JJs}b shows conductance $G$ versus temperature $T$ curves at two different biases ($V=0.80~\text{mV} ~\text{and}~0.31~\text{mV}$). A sudden jump in conductance at $T\sim5$~K marks the superconducting transition of the vanadium electrodes. For both applied biases, $G$ takes the same values in the 2~K to 6~K temperature range. Although in both cases the same conductance level is reached at $T=0.3~\text{K}$, with $V=0.80~\text{mV}$ this $G$ level is only reached at lower temperatures than with $V=0.31~\text{mV}$. The situation becomes more clear with the differential conductance vs bias curves presented in Figure~\ref{fig-JJs}c. The two different levels of conductance below and above the critical temperature of V are clearly seen by comparing the $T=3.2$~K and $T=8.0$~K curves. Finally, it can be seen that the increase in conductance seen below $T=2~K$ in Figure~\ref{fig-JJs}b corresponds to a step-like increment in conductance in the $T=0.3$~K $G(V)$ curve at $\vert{V}\vert\lesssim1.5$~ mV. It becomes apparent that this low bias increase in conductance slowly emerges below $T=2$~K. 
This behavior with an additional low bias excess conductance of the bridge emerging below 2~K is in agreement with previous observations of the temperature range in which vanadium-based JJs become operational~\cite{Ligato2017}.

We performed transport measurements while applying a rotating in-plane magnetic field to see the influence of changing the orientation of the magnetization of the Fe bridge on low bias conductance. Figure~\ref{fig-JJs}d shows $G$ as a function of the direction $\theta$ of the applied in-plane magnetic field $H$ for different applied current densities across the bridge. There are two well-defined high and low conductance levels with sudden transitions and some intermediate states. In our view, these conductance level transitions occur when the Fe layer changes its magnetization direction due to the action of the magnetic field. The results are compatible with theoretical predictions of the spin supercurrent dependence on the magnetization direction of the ferromagnetic weak link in lateral JJs~\cite{Bujnowski2019,eskiltprb2019}. 

We tentatively attribute the effects of varying the current density to either the influence of Josephson current on magnetic exchange interactions \cite{Waintal2002}, or to the influence of spin supercurrent exerting a torque on the magnetization of the ferromagnetic Fe(100) layer~\cite{Linder2014}.
In that context, increasing the spin supercurrent density could influence the modified IP MCA energy of the Fe(100) layer (Section~\ref{section:IP-MCA}), creating a new MCA energy minimum at $\theta=135$\degree. This would explain the 135\degree switch of the magnetization orientation observed with $J=2.5\times10^8\text{A/m}^2$ compared to the switch at 90\degree (suppressing Josephson current) when the current density exceeds $J=4.2\times10^8\text{A/m}^2$. We have evaluated the possible influence of the current induced magnetic field on the above observed effects. Our COMSOL calculations yield a maxium self-induced magnetic field of about $\sim0.5$~mT, which in our view is too small compared to the rotated external magnetic field of $30$~mT and therefore should not produce any substantial impact on the magnetic reorientations.

Finally, Figure~\ref{fig-JJs}e shows $G$ vs $T$ experiments under different orientations of the applied in-plane magnetic field. Here, the conductance is normalized so that $G^{N}(\theta,T)=G(\theta,T)/G(\theta,T=4\text{~K})$ to facilitate the comparison between the different curves. For all $\theta$ there is a marked and reproducible contact superconducting transition at $T=5$~K, however, the characteristic conductance increase under 2~K varies. The inset shows the behavior of $G(\theta,T=0.3~\text{K})$ as a function of $\theta$, which is also clearly anisotropic. The difference in the anisotropic response of the excess conductance when compared to the rotation experiments shown in panel~(d) could be due to the following reasons. First, in panel~(d) the magnetization is rotated at low temperatures, while in the experiment shown in panel~(e) the magnetization direction is established much above $T_C$ ($T=16$~K). Secondly, the conductance shown in panel~(c) is differential conductance, measured at specific current densities, while panel~(e) shows the \textit{integrated conductance}, determined by the inverse resistance of the junction. As shown in panel~(c), the excess-low bias conductance at temperatures below 2~K depends on the applied current density, and thus the angular dependence of the integrated response in panel~(e) should be different.

To sum up this section, we have presented promising results which point towards the possible presence of LRTs in our lateral Josephson junctions. We do not completely exclude the possible presence of pinholes in these junctions. Further detailed research is planned to fully understand the above-described phenomena and gather conclusive evidence on the possibility of controlling spin supercurrents via the magnetization orientation in lateral S/F/S JJs with interfacial SOC. Interestingly, theoretical predictions reveal a nonmonotonic spin-triplet contribution with the strength of the interfacial barrier and spin-orbit coupling~\cite{Shen2024}.

\section{Outlook}\label{section:Outlook}

The structure of this manuscript reflects the fact that Fe/MgO/V-based heterostructures are currently the only systems in which superconducting spintronics involves the interplay of Cooper pairing, electron symmetry filtering, and interfacial spin-orbit coupling. Similar studies  would be of interest in superconductor-antiferromagnet (S/AF) juntions with SOC. Indeed, recently, attention has turned to S/AF hybrids. Although most of this works are theoretical, they suggest that S/AF systems, despite having nearly zero net magnetization, could also host non-trivial spin-triplet correlations when SOC or non-collinear spin textures are present. Notably, the mechanisms generating long-range triplet Cooper pairs in these systems differ from those in conventional S/F structures, as they do not rely on exchange splitting of otherwise degenerate energy levels. Jakobsen and co-workers~\cite{Jakobsen2020} theoretically examined electrical and thermal transport in S/AF junctions with Rashba SOC. They predict that rotating the Néel vector from an in-plane orientation to one perpendicular to the interface can produce a large low-bias anisotropic magnetoresistance, qualitatively similar to the magneto-anisotropic Andreev reflection effect discussed earlier for S/F systems with Rashba SOC (see Chapter~\ref{section:MAAR} and Ref.\cite{Martinez2020}).

In agreement with these findings, Bobkov et al. report that Rashba SOC can anisotropically modify proximity-induced spin-triplet correlations, leading to an orientation-dependent modulation of the superconducting critical temperature as the Néel vector is rotated relative to the S/AF interface~\cite{Bobkov2023}.
To account for a stronger-than-expected suppression of the critical temperature in S/AF bilayers, another mechanism that does not involve SOC has been proposed. In compensated antiferromagnets, the superconducting proximity effect can generate Néel-triplet Cooper pairs whose pairing amplitude oscillates spatially in step with the AF lattice spin structure~\cite{Bobkov2022}. While this mechanism weakens superconductivity in clean systems, interfacial disorder that induces a local ferromagnetic moment offers an alternative route for generating spin-triplet correlations in S/AF junctions.
Fyhn et al. predict that in disordered metallic antiferromagnets, impurities (especially those near the interface) that promote interfacial ferromagnetism, suppress singlet correlations but leave triplet correlations largely unaffected~\cite{Fyhn2023}. Adjusting the direction of this interfacial ferromagnetic moment relative to the bulk Néel vector can selectively filter the spin-triplet component of the proximity effect, providing a potential means to realize a magnetic switch for spin supercurrents in S/AF junctions.

In this work, we have focused on the anomalous electron transport in S/F junctions with SOC, leaving aside several other important effects that have recently emerged in S/F hybrids with SOC as they are not directly related to epitaxy and, therefore, orbital symmetry filtering. One such effect is ferromagnetic resonance (FMR)-induced spin supercurrent pumping~\cite{Yao2018,Jeon2020}. Our currently ongoing dynamic experiments aim to investigate the temperature dependence of damping in V/MgO/Fe-based junctions. Preliminary results suggest an enhanced FMR damping in the superconducting state compared to the normal state, which could provide independent evidence of LRT generation induced by SOC localized at the S/F interface (in contrast to previous reports where SOC was localized aside the S and F layers~\cite{Jeon2020}). Additionally, we have not discussed the magnetic control of giant thermoelectric effects in V/MgO/Fe based junctions~\cite{ouassou-prb-22,GonzlezRuano2023, Tuero2025}, since both SOC and symmetry filtering are not critical for their observation. Recent theoretical work, however, predicts that the symmetry of the thermoelectric response in S/F junctions with SOC could serve as yet another probe of LRT generation~\cite{Dutta2020}.

Finally, we believe that the above discussed effects could be explored in other epitaxial heterostructures such as V/MgO/Co or Nb/MgO/Fe - based systems. In the junctions involving Co instead of Fe, the thicknes of Co layer should be only of few nanometers. This is because ultrathin Co grown on top of MgO could transform its hexagonal crystalline structure to a body-centered one, and provide even better orbital symmetry and spin filtering effects than the one taking place in V/MgO/Fe~\cite{Yuasa2006}.
On the other hand, replacing vanadium with bcc Niobium could probably not be feasible for superconducting spintronics with electron symmetry filtering and interfacial spin-orbit coupling. First, the lattice mismatch between superconductor and MgO would increase by roughly a factor of three~\cite{Guasco2023}, thereby degrading the epitaxial growth and enhancing the concentration of extended defects such as dislocations with a significant effect on noise. Secondly, as niobium has pronounced $\Delta_1$ states at the Fermi level~\cite{Mattheiss1970}, the impact of SOC on electron and spin transport in Nb/MgO/Fe heterostructures would be greatly reduced due to an additional shunting current channel between the electrodes created by the $\Delta_1$ “hot spots” in MgO.

\section{Summary and conclusions}\label{section:Conclusions}

The main goal of this manuscript has been to explore the mutual interplay between magnetism and superconductivity in fully epitaxial V/MgO/Fe heterostructures. Importantly, electronic transport in these superconducting spintronic devices is governed by spin and symmetry filtering processes mediated by interfacial SOC at the V/MgO and MgO/Fe interfaces. At room temperature, we have observed an increase in the TMR under applied bias, in contrast to the TMR suppression with bias typically seen in standard  MTJs. This result highlights the crucial role of SOC and extends the applicability of spintronic devices towards higher operational biases. In the superconducting state, and in line with theoretical predictions, we have gathered several indirect but consistent experimental signatures of controlled equal-spin triplet Cooper pair generation. These include: a three-orders-of-magnitude enhancement in zero-bias conductance anisotropy, superconductivity-induced modifications of the magnetic anisotropy and magnetization-sensitive above-gap conductance anomalies (MacMillan resonances). MacMillan resonances allowed us to separate the SOC and spin-texture related contributions to equal spin triplet generation. Together, these findings point to orbital symmetry and SOC-controlled spin filtering as a promising platform for advancing superconducting spintronics.

Furthermore, very recent shot noise experiments on the same type of junctions have revealed an excess -- of nearly an order of magnitude -- below the superconducting gap compared to expected values. This may constitute additional evidence of long-range triplet formation due to SOC. Finally, our preliminary (ongoing) studies on laterally patterned JJs based on V/MgO/Fe-based heterostructures open the possibility of obtaining more direct evidence for the role of triplet Cooper pairs, through the investigation of the dependence of the Josephson effect on the in-plane orientation of the Fe(100) magnetization. Overall, our results establish epitaxial V/MgO/Fe heterostructures as strong candidates for building blocks of the next-generation superconducting spintronic devices, with potential applications ranging from ultra–low-energy cryogenic memories to quantum computing architectures.

\section*{Data availability statement}

The data that support the findings of this study are available upon reasonable request from the authors.

\section*{Acknowledgements}

We thank Diego Caso, Isidoro Mart\'inez and Juan Pedro Cascales for their help with experiments and simulations. Jaroslav Fabian, Jacob Linder, Jong Han, Chenghao Shen, Michel Hehn, Pilar Prieto, Andreas Costa and Alex Matos-Abiague are also acknowledged for the fruitful discussions and collaboration. Finally, Manuel Rodr\'iguez and Fernando Jim\'enez from IMDEA Nanoscience are acknowledged for their help with fabrication of lateral Josephson junctions. 
The work in Madrid was supported by Spanish Ministry of Science and Innovation (PID2021-124585NB-C32, TED2021-130196B-C22 and PID2024-155399NB-I00). F.G.A. also acknowledges financial support from the Spanish Ministry of Science and Innovation through the Mar\'ia de Maeztu Programme for Units of Excellence in R\&D (CEX2023-001316-M). This research also received support from the Comunidad de Madrid through project TEC-2024/TEC-380 (Mag4TIC-CM). The growth of samples was performed using equipment from the CC-DAUM platform funded by FEDER (EU),ANR, the Region Lorraine, and the metropole of Grand Nancy. I.\v{Z}. acknowledges U.S. DOE, Office of Science BES, Award No. DE-SC0004890. C.T. acknowledges the funding project UEFISCDI via the project “MODESKY” ID PN-III-P4-ID-PCE-2020-0230-P, Grant No. UEFISCDI: PCE 245/02.11.2021.

%\printbibliography

%\end{document}
%\bibliographystyle{unsrt}
%\bibliographystyle{apsrev4-1}
\bibliography{REFERENCIAS_BUENAS}

\end{document}